\documentclass[aps,pra,reprint]{revtex4-2}
\usepackage{amsmath}

\usepackage{graphicx}
\graphicspath{{figures/}}
\usepackage{amsfonts}
\usepackage{color}
\usepackage{mathtools}

\usepackage{hyperref}
\usepackage[capitalize]{cleveref}
\usepackage{verbatim} 
\hypersetup{
	colorlinks = true,
	linkcolor =blue,
	citecolor=blue, 
	urlcolor=blue 
}

\usepackage{braket}
\usepackage{physics}
\usepackage{bm}
\usepackage{amsthm}

\newcommand{\bobkey}{Z^n_B}

\newcommand{\epsAT}{\varepsilon_\text{AT}}
\newcommand{\epsPA}{\varepsilon_{\text{PA}}}
\newcommand{\epsbar}{\bar{\varepsilon}}
\newcommand{\epsEV}{\varepsilon_{\text{EV}}}
\newcommand{\epsSec}{\varepsilon_{\text{secure}}}
\newcommand{\epscorrect}{\varepsilon_1}
\newcommand{\epssecret}{\varepsilon_2}

\newcommand{\half}{\frac{1}{2}}
\newcommand{\Cfull}{\widetilde{C}}
\newcommand{\Cbar}{\bar{C}}
\newcommand{\Cat}{C_{AT}}

	\newcommand{\floor}[1]{\lfloor #1 \rfloor}
	\newcommand{\ceil}[1]{\lceil #1 \rceil}
	\newcommand{\tracenorm}[1]{\biggl|\!\biggl| #1 \biggr|\!\biggr|_1}

\newcommand{\accset}[1]{\widetilde{Q}_{#1}}
\newcommand{\accevent}[1]{\widetilde{\Omega}_{#1}}
\newcommand{\acceventEV}{\Omega_{\text{EV}}}
\newcommand{\fullaccevent}[1]{\accevent{#1} \! \wedge \! \acceventEV}
\newcommand{\Sset}[1]{\widetilde{S}_{#1} }

\newcommand{\acceventadapt}[1]{\Omega_{#1}}
\newcommand{\accsetadapt}[1]{Q_{#1}}
\newcommand{\fullacceventadapt}[1]{\Omega_{#1}  \! \wedge \! \Omega_\text{{EV}}} 

\newcommand{\finalstatefull}[1]{ \rho^{(l_{#1}, \leak{#1} )}_{K_A K_B \Cfull E^N| \fullaccevent{#1}  } }
\newcommand{\finalstatefullideal}[1]{ \rho^{(l_{#1}, \leak{#1},\text{ideal} )}_{K_A K_B \Cfull E^N| \fullaccevent{#1}  } }
		
\newcommand{\finalstate}[1]{ \rho^{(l_{#1}, \leak{#1} )}_{K_A \Cfull E^N| \fullaccevent{#1}  }}
\newcommand{\finalstateideal}[1]{ \rho^{(l_{#1}, \leak{#1},\text{ideal} )}_{K_A  \Cfull E^N| \fullaccevent{#1}  }}

\newcommand{\finalstatefulladapt}[1]{ \rho^{(l_{#1}, \leak{#1} )}_{K_A K_B \Cfull E^N| \fullacceventadapt{#1}  } }

\newcommand{\finalstateadapt}[1]{ \rho^{(l_{#1}, \leak{#1} )}_{K_A \Cfull E^N| \fullacceventadapt{#1}  }}
\newcommand{\finalstateidealadapt}[1]{ \rho^{(l_{#1}, \leak{#1} ,\text{ideal})}_{K_A  \Cfull E^N| \fullacceventadapt{#1}  }}

\newcommand{\smoothmin}{H^{\epsbar}_{\text{min}}}
\newcommand{\renyi}{R\'{e}nyi }

\newcommand{\EVCost}{\ceil{ \log \left( 1/\epsEV \right) }}

\newcommand{\vecb}[1]{\mathbf{#1}} 

\newcommand{\Fobs}{\vecb{F}^{\text{obs}}}
\newcommand{\Fbar}{\bar {\vecb{F}}}

\newcommand{\bstat}{b_\text{stat}}
\newcommand{\leak}[1]{\lambda^{\text{EC}}_{#1}}

\newcommand{\nch}{n_\text{ch}}

\newcommand{\Zbot}{\hat{Z}}
\newcommand{\Zv}{{Z^{\leq n}}}

\newcommand{\zvone}{z_1}
\newcommand{\zvtwo}{z_2}

\newcommand{\hashfamily}[1]{\mathcal{F}^{\text{hash}}_{#1}}
\newcommand{\virthashfamily}[1]{\hat{\mathcal{F}}^{\text{hash}}_{#1}}

\newcommand{\KRfixed}[1]{R_\text{fixed,#1}}
\newcommand{\KRvariable}[1]{R_\text{variable,#1}}

\newcommand{\PAcost}{  \theta(\epsPA,\epsEV)}

\newcommand{\honest}{\rho_\text{hon}}

\theoremstyle{definition} 
\newtheorem{theorem}{Theorem}
\newtheorem{definition}{Definition}
\newtheorem{lemma}{Lemma}

\newtheorem{remark}{Remark}

\begin{document}
	\title{Security Proof for Variable-Length Quantum Key Distribution}  
	\author{Devashish Tupkary}
	\email{djtupkary@uwaterloo.ca}
	\affiliation{Institute for Quantum Computing and Department of Physics and Astronomy, University of Waterloo, Waterloo, Ontario, Canada, N2L 3G1}	
	
		\author{Ernest Y.-Z. Tan}
		\email{yzetan@uwaterloo.ca}
	\affiliation{Institute for Quantum Computing and Department of Physics and Astronomy, University of Waterloo, Waterloo, Ontario, Canada, N2L 3G1}	
	
	\author{Norbert L\"utkenhaus}
	\email{nlutkenhaus.office@uwaterloo.ca}
	\affiliation{Institute for Quantum Computing and Department of Physics and Astronomy, University of Waterloo, Waterloo, Ontario, Canada, N2L 3G1}

	\begin{abstract}
		We present a security proof for variable-length QKD in the Renner framework against IID collective attacks. Our proof can be lifted to coherent attacks using the postselection technique. Our first main result is a theorem to convert a sequence of security proofs for fixed-length protocols satisfying certain conditions to a security proof for a variable-length protocol. This conversion requires no new calculations, does not require any changes to the final key lengths or the amount of error-correction information, and at most doubles the  security parameter.  Our second main result is the description and security proof of a more general class of variable-length QKD protocols, which does not require characterizing the honest behaviour of the channel connecting the users before the execution of the QKD protocol. Instead, these protocols adaptively determine the length of the final key, and the amount of information to be used for error-correction, based upon the observations made during the protocol. We apply these results to the qubit BB84 protocol, and show that variable-length implementations lead to higher expected key rates than the fixed-length implementations.
	\end{abstract}
	
	\maketitle
	
	\section{Introduction} \label{sec:introduction}

 Security proofs for QKD protocols are typically proven in the ``fixed-length'' scenario, where two users Alice and Bob either produce a key of a fixed length, or abort the protocol \cite{tomamichelLargely2017,georgeNumerical2021,rennerSecurity2005,bunandarNumerical2020,ruscaFinitekey2018,limConcise2014}. Such protocols accept and produce a key of fixed length if and only if their observed statistics belong to some predetermined ``acceptance set".  Otherwise, the protocol aborts.  Such protocols have two main disadvantages.
 
First, in order to ensure that the protocol accepts with high probability for honest behaviour, the acceptance set needs to be chosen carefully. Typically, the acceptance set is chosen to be the set of statistics that are close to what is expected from honest behaviour \cite{georgeNumerical2021,bunandarNumerical2020,rennerSecurity2005}. This requires Alice and Bob to know the honest behaviour of the channel connecting Alice and Bob, \emph{before} a run of the QKD protocol. In many practical scenarios, such as ground-to-satellite QKD \cite{bourgoinComprehensive2013,dequalFeasibility2021,liaoSatellitetoground2017,trinhStatistical2022,sidhuFinite2022}, it is difficult to know the behaviour of the channel in advance. In fact, this can be a problem even in fibre-based setups \cite{Wang2022,Clivati2022,Dynes2012}. 

Second, even if the honest behaviour is known, the size of the acceptance set affects the length of the final key that can be produced. This reflects the fact that the key has to be secure for the \textit{worst-case} event that accepts. Larger acceptance sets have a high probability of accepting on any given run of the QKD protocol, but lead to a shorter length of the final key, since they include worse accept events. In particular, if users choose a large acceptance set, and then find that their observed statistics are much better than expected, they are \emph{not} allowed to produce a larger key.
Thus, there is a trade-off between protocols that accept with high probability, and protocols which produce a large key on accepting.

A variable-length QKD  protocol is one that allows users to adjust the length of the key generated based upon the observed statistics during the protocol \cite{ben-orUniversal2004,hayashiConcise2012}.  This eliminates the trade-off described above.  It also does not require the expected behaviour of the channel to be known in advance. In fact, many prior works have implemented such a variable-length protocol based upon intuition. For the qubit BB84 protocol, a rigorous treatment of variable-length protocols can be found in Ref.~\cite{hayashiConcise2012}, using the phase-error approach for security proofs.

In this work, we present a security proof for variable-length QKD protocols against IID collective attacks, which can be lifted to coherent attacks using the postselection technique \cite{christandlPostselection2009}. Since that lift to coherent attacks is technical and requires details of the postselection technique, it is included in Ref.~\cite{nahar2024postselection} for pedagogical reasons. This work differs from Ref.~\cite{hayashiConcise2012} in that it follows the lines of the Renner framework (i.e.~bounding suitable entropies and applying a leftover hashing lemma) rather than the phase-error approach; in particular, it does not involve an explicit reduction to an analysis of a virtual phase-error-correction procedure. 
 
 This work is organized as follows: In \cref{sec:notation} we describe the QKD protocol steps, and setup the notation used in this work. In \cref{sec:fixedtoadaptive} we show how, under certain conditions,  a sequence of fixed-length QKD security proofs against IID collective attacks can be lifted to a variable-length QKD security proof against IID collective attacks. Our approach involves at most a doubling of the security parameter. Therefore, for the same target security parameter, the various key lengths for the variable-length protocol are nearly identical to the key lengths for the fixed-length protocols. In \cref{sec:applicationbb84}, we consider the scenario where the expected channel behaviour is known in advance, and compute expected key rates for the qubit BB84 protocol, and show that the variable-length protocol generates better expected key rates than the best fixed-length protocol. 
		
	We then move on to study scenarios where the channel behaviour is unpredictable, and \textit{not known} in advance. 	In \cref{sec:truevariablelength} we present another variable-length protocol, where the procedure for choosing the final key length (and length of error-correction information) is especially suited for such scenarios. Our protocol allows Alice and Bob to perform QKD without any prior knowledge about the channel connecting them. In \cref{sec:resultstrueadaptive} we apply these results to the qubit BB84 protocol, and show that the variable-length implementation lead to higher expected key rates than the best fixed-length implementation.
		
	 In \cref{sec:variablelengthPA}, we point out and remedy a gap between the theory and implementation of  privacy amplification in QKD protocols. This gap exists because in implementations, privacy amplification is typically done on a register of variable length containing the raw sifted key, whereas in theory, privacy amplification is typically done on a register of fixed length containing data from all the signals. In \cref{sec:conclusion} we present concluding remarks. Various technical details are delegated to the appendices.

	\section{Notation and Protocol Specifications} \label{sec:notation}
	In this work, we will either consider a \textit{sequence of fixed-length} protocols indexed by $i$, or a \textit{single variable-length} protocol where different events in the variable-length decision (see below) are indexed by $i$. We use the same index $i$ since we construct the variable-length protocol from the sequence of fixed-length protocols in \cref{sec:fixedtoadaptive}. We describe the protocol steps for both fixed-length and variable-length protocols below.
	
	\subsection{Protocol Steps}
	
		\begin{enumerate}
		\item \textbf{State Preparation and Transmission:} Alice prepares signal states and sends them to Bob, who measures them. We let $N$ be the total number of signals sent by Alice. 
		For prepare-and-measure protocols, the source-replacement scheme \cite{curtyEntanglement2004} can be used to equivalently describe this step as Alice and Bob receiving subsystems of the state $\rho_{A^N B^NE^N} $, followed by Alice measuring her subsystem $A^N$. In this case, one can assume that $\Tr_{B^NE^N}(\rho_{A^NB^NE^N}) = \bar{\sigma}_A^{\otimes N}$, where $\bar{\sigma}_A^{\otimes N}$ is a fixed marginal state which reflects the fact that Alice's system never leaves her lab, and that each signal is prepared independently. Furthermore, since we assume IID collective attacks, we have $\rho_{A^N B^NE^N} = \rho^{\otimes N}_{ABE}$.
		\item \textbf{Measurement:} Bob performs measurements on the received states, and stores measurement data. 
		\item \textbf{Public Announcements: } Alice and Bob select at random $m$ rounds~\footnote{For the purposes of this work we take $m$ to be a constant; with minor modifications our proof should generalize to the case where $m$ could be a random variable.} out of the total $N$ rounds, and announce their measurement outcomes for those rounds in the register $\Cat^m$. These announcements will be used to determine whether to accept or abort in the fixed-length protocol, or to determine appropriate lengths of various strings in the variable-length protocol.
		
		On the remaining $n\coloneqq N-m$ 
		rounds, Alice and Bob perform round-by-round announcements  $C^n$ (such as basis-choice, detect / no-detect). They store their private data in registers $X^n$ and $Y^n$. The state of the protocol at this stage is given by $\rho_{X^nY^nC^n \Cat^m E^N} = \rho^{\otimes n}_{XYCE} \otimes \rho_{\Cat^m  E^{m}}$, where we split up the $m$ test rounds and $n$ key generation rounds. Note that since we assume IID collective attacks, the test round announcements $\Cat^m$ and registers $E^m$ are independent of the raw key $Z^n$.
		
		 From the public announcements $\Cat^m$, Alice and Bob compute $\Fobs$, which is the observed frequency of outcomes in the test rounds of the QKD protocol.
		\item \textbf{Acceptance Test / Variable-Length Decision: }  For the $i$th \textit{fixed-length} protocol, Alice and Bob accept the protocol if $\Fobs \in \accset{i}$. Here $\accset{i}$ denotes the acceptance set for the protocol, and we use $\accevent{i}$ to denote the event $\Fobs \in \accset{i}$.  Note that in this work, we use variables $\Omega$ and $\widetilde{\Omega}$ (with subscripts) to denote the boolean variables corresponding to the occurrence of various events.
		
		 For the \textit{variable-length} protocol, we instead use the following procedure: we have multiple disjoint sets $\accsetadapt{i}$, and use $\Omega_i$ to denote the event $\Fobs \in \accsetadapt{i}$. Depending on which event $\Omega_i$ is observed, Alice and Bob can choose different parameters in the processing of the data to the final key (for instance, the number of bits used in error correction, and the length of the final key).
		\begin{remark} \label{rem:acceptancetest}
			It is important to note that for fixed-length protocols, the details of the acceptance test need to be determined \textit{before} looking at $\Fobs$. In particular, current security proofs for such protocols do not allow users to \textit{first} look at the observed statistics $\Fobs$ and \textit{then} decide the nature of the acceptance test. This is the reason why it is important to know the expected behaviour of the channel \textit{before} the QKD protocol is run, in order to design an acceptance test that accepts with high probability for honest behaviour.
		\end{remark}
		
		\item \textbf{Key Map and Sifting: }Alice maps her raw data $X^n$ to her raw key $Z^n$ where $Z$ is a binary variable, based on the announcements $C^n$. In this work, we assume that Alice sets $Z$ to $0$ for signals that are sifted out, and let $d_Z$ denote the dimension of the $Z$ register.
		
		\item \textbf{Error-correction: }Alice and Bob implement error-correction by exchanging classical information in the register $C_E$. For the $i$th \textit{fixed-length} protocol, we use $\leak{i}$ to denote the number of bits of communication during error-correction, when the protocol accepts. For \textit{variable-length} protocols we use $\leak{i}$ to denote the number of bits of communication during error-correction, when event $\Omega_i$ occurs.  Note that $C_E$ may contain additional information beyond $\leak{i}$ bits, as long as  the information is independent of Alice and Bob's data. For example, if the error-correction protocol randomly divides the data into blocks, then the descriptions of the randomly generated blocks can be included in $C_E$. Thus $\leak{i}$ actually refers to the number of bits in $C_E$ that are computed from  Alice and Bob's data.
		
		\begin{remark}
			It is important to note that one has to fix the exact number of bits of communication during error-correction \textit{before} the QKD protocol is run. In particular, current security proofs do not allow users to \textit{first} count the number of bits used during error-correction  and \textit{then} adjust the length of the final key produced. For the framework described in this work for variable-length protocols, it is still the case that the values $\leak{i}$ need to be decided before the protocol is run, i.e.~for each $i$, the users must implement an error-correction procedure that uses a fixed number~\footnote{This condition can be slightly weakened to having a fixed upper bound $\nu^{\text{EC}}_{i}$ on the number of bits used, by noting that the number of bitstrings of length \textit{up to} some value $\nu$ is $2^{\nu+1}-1$, so an $(\nu+1)$-bit register suffices to encode all such bitstrings. With this, it suffices to replace the $\leak{i}$ values in our subsequent key length formulas with $\nu^{\text{EC}}_{i}+1$.} of bits, rather than one that uses a randomly varying number of bits.
		\end{remark}
	
		\item \textbf{Error-verification: }Alice chooses a two-universal hash function that hashes to $ \ceil {\log(1/\epsEV)}$ bits, computes the hash of her raw key, and sends the hash value to Bob, along with the description of the hash function. Bob hashes his guess for Alice's key, compares the hash values, and announces whether the values match or not. We use $C_V$ to denote the classical register that stores this communication. We note that since the hash function is chosen independently of Alice and Bob's data, only $ \ceil{\log(1/\epsEV)}$ bits of $C_V$ are correlated to Alice and Bob's data \footnote{While Bob's announcement technically constitutes an extra bit, we note that in our security proofs, when accounting for the ``leakage'' caused by $C_V$, we only consider the state conditioned on accepting in this step, in which case this extra bit takes a deterministic value and does not affect any entropies.}.  We use $\acceventEV$ to denote the event that the hash values match, and Alice and Bob continue with the protocol. 
		The state of the protocol at this stage is given by $\rho_{Z^n \bobkey C^n  C_E C_V E^n } \otimes \rho_{ \Cat^m E^m}$, where $\bobkey$ is Bob's guess for Alice's raw key after error-correction.

		\item \textbf{Privacy Amplification:}		For the $i$th  \textit{fixed-length} protocol, if event $\accevent{i} \wedge \acceventEV$ occurs, Alice chooses a two-universal hash function from $n$ bits to $l_i$ bits. She announces the description of the hash function in the register $C_P$, and Alice and Bob apply the hash function to their data to produce their final keys in registers $K_A$ and $K_B$.  We use $\finalstatefull{i}$ to denote the final state of the protocol, conditioned on the event $\accevent{i} \wedge \acceventEV$, where we use $\Cfull$ to denote the registers $C^n \Cat^m C_EC_VC_P$ for brevity.
		
	 For the \textit{ variable-length} protocol, if event $\fullacceventadapt{i} $ occurs, Alice chooses a two-universal hash function from $n$ bits to $l_i$ bits. She announces the description of the hash function in the register $C_P$, and Alice and Bob apply the hash function to their data to produce their final keys in registers $K_A$ and $K_B$.  We use $\finalstatefulladapt{i}$ to denote the final state of the protocol, conditioned on the event $\fullacceventadapt{i}$. 
	 
	 	\end{enumerate}

		Thus for all the  fixed-length QKD protocols, the details of the acceptance test, and the value of $\leak{i}$ and $l_i$ must be fixed \textit{before} the start of the protocol. For the variable-length QKD protocol, the details of the variable-length decision (in particular the values of $\leak{i}$ and $l_i$) must be fixed \textit{before} the start of the protocol. Moreover, the events $\accevent{i}$,$\acceventadapt{i}$ determine the pair of values $(l_i,\leak{i})$ for the corresponding protocol.

	\section{Variable-Length Security from Fixed-Length Security}  \label{sec:fixedtoadaptive}
	In this section, we show how a sequence of security proofs for fixed-length protocols (against IID collective attacks) can be converted to a security proof for variable-length protocols (against IID collective attacks).  Let us suppose that we have $M$ fixed-length QKD protocols, indexed by $i \in \{1,2,\dots,M\}$. The protocols differ only in their choice of acceptance test, the number of bits used for error-correction, and length of the final key generated. In particular, the $i$th protocol accepts if and only if $\Fobs \in \accset{i}$, where $\accset{i}$ is the acceptance set. Upon acceptance, it uses $\leak{i}$ bits for error-correction and produces a key of fixed length $l_i$. 
	\subsection{Fixed-Length Security Statements} \label{subsec:fixedlengthstatements}

Following standard composable security definitions~\cite{portmannCryptographic2014a,portmannSecurity2022}, the $i$th fixed-length QKD protocol is said to be $\epsSec$-secure against some class of attacks if the following condition holds: for all attacks in that class, at the end of the protocol we have
		\begin{equation} \label{eq:securitystatement}
			\begin{aligned}
		\half \Pr( \fullaccevent{i} \! )  &\norm{ \finalstatefull{i} \!-  \finalstatefullideal{i} }_1 \\
		 &\leq \epsSec.
		\end{aligned}
	\end{equation}
Here, $\finalstatefull{i}$ denotes the actual output state at the end of the protocol, while $\finalstatefullideal{i}$ denotes an ``ideal state'' obtained by replacing the key registers of $\finalstatefull{i}$ with perfect keys, i.e.
	\begin{equation} \label{eq:idealfixedlength}
		\begin{aligned}
		\finalstatefullideal{i} \coloneqq \sum_{k \in \{0,1\}^{l_{i}}} \frac{\ket{kk} \bra{kk}
		}{2^{l_{i}}} \otimes \rho^{(l_{i}, \leak{i} )}_{\Cfull E^N| \fullaccevent{i}  } .
		\end{aligned}
	\end{equation}
	Note that $\finalstatefullideal{i}$ is not a ``fixed'' state but rather a function of the input state. 
	
	In particular, in this work we focus on restricting the class of attacks to IID collective attacks, which means we suppose that the input states supplied to the QKD protocol are always of the form $\rho^{\otimes N}_{ABE}$. For prepare-and-measure protocols, the input states are further constrained to satisfy $\Tr_{BE} (\rho_{ABE}) = \bar{\sigma}_A$, where $\bar{\sigma}_A$ is the fixed marginal state on Alice's system that is obtained from the source-replacement scheme \cite{curtyEntanglement2004}. (For entanglement-based protocols, this constraint is not imposed; either version can be handled using the framework presented in this work.)

Furthermore, as explained in Ref.~\cite{portmannCryptographic2014a,portmannSecurity2022,ben-orUniversal2004}, to show that a QKD protocol is $\epsSec$-secure (against some class of attacks), it suffices to prove a pair of simpler conditions, namely that it is $\epscorrect$-correct and $\epssecret$-secret (against that class of attacks) with $\epscorrect + \epssecret \leq \epsSec$. Specifically, $\epscorrect$-correctness means the output state satisfies
\begin{equation} \label{eq:corrdef}
	\begin{aligned}
			\Pr(K_A \neq K_B \wedge \acceventEV) \leq \epscorrect,
	\end{aligned}
\end{equation}
while $\epssecret$-secrecy means it satisfies 
\begin{equation}
\begin{aligned}
\frac{1}{2}	\Pr( \fullaccevent{i} ) &\tracenorm{ \finalstate{i}  - \finalstateideal{i}} \\
&\leq \epssecret,
\end{aligned}
\end{equation}
i.e.~the same condition as $\epsSec$-security (\cref{eq:securitystatement}) but with Bob's key register omitted. (Both of the above conditions are to be implicitly understood as holding against all attacks in the considered class.) In our subsequent discussion, we shall indeed proceed by proving the above pair of conditions rather than \cref{eq:securitystatement} directly.

We assume that for each protocol $i \in \{1,2,\dots,M\}$, the following statements have been shown to be true. As we shall shortly show, these statements together imply \cref{eq:securitystatement} with $\epsSec = \epsEV + \max\{\epsAT,\epsPA\}$, following established approaches 
described in e.g.~\cite{georgeNumerical2021,rennerSecurity2005,bunandarNumerical2020}.
	\begin{enumerate}
		\item There is a ``feasible'' set $\widetilde{S}_i \subseteq \{ \rho \in  S_\circ(AB) | \Tr_{B} (\rho_{AB} ) = \bar{\sigma}_A \}$, where $S_\circ$ denotes the set of normalized states, such that if the state $\rho_{AB}$ is not in the set $\widetilde{S}_i$, the protocol aborts with high probability, and is thus secure. That is, 
		
		\begin{equation} \label{eq:Sdef}
			\rho_{AB} \notin \Sset{i} \implies \text{Pr}\left(\widetilde{\Omega}_i \right) \leq \epsAT.
		\end{equation}
		Note that in the entirety of this work, the statement $\rho \notin \Sset{i}$ is assumed to be with respect to the parent set $ \{ \rho \in  S_\circ(AB) | \Tr_{B} (\rho_{AB} ) = \bar{\sigma}_A \}$ having the fixed marginal on $A$. 
	
		\item The hash length $l_i$ is given by
			\begin{equation} \label{eq:l_ivalue}
			\begin{aligned} 
				l_i = &\left \lfloor n \min_{\rho \in \widetilde{S}_i} H(Z | C E)_{\rho} - \leak{i} - \EVCost \right. \\
				& \left. - n (\alpha-1) \log^2(d_Z+1)  . - \frac{\alpha}{\alpha-1} \left( \log( \frac{1}{4 \epsilon_\text{PA} })  + \frac{2}{\alpha} \right)  \right \rfloor,
			\end{aligned}
		\end{equation}
		where $H$ denotes the conditional von Neumann entropy, and $d_Z$ denotes the dimension of the $Z$ register. This choice of $l_i$ is such that if the state $\rho_{AB} \in \Sset{i}$, a key of length $l_i$ can be safely extracted from the protocol. 	In the entirety of this work, we choose $\alpha = 1 + \kappa/\sqrt{n}$ with $\kappa \coloneqq \sqrt{{\log(1/\epsilon_\text{PA})}}/{\log(d_Z+1)}$, assuming $n$ is large enough to ensure that $\alpha \leq 1+ 1/\log(2d_Z+1)$ is satisfied. This is the choice of $\alpha$ that maximizes \cref{eq:l_ivalue} (up to a minor approximation that $\alpha/(\alpha-1) \approx 1/(\alpha-1)$), 
		and also leads to the expected asymptotic scaling in the key rate expression.
		
		\item The error-verification step compares two-universal hashes of length $\ceil{\log(1/\epsEV)}$. 
		
	\end{enumerate}
		
To see how these three statements imply \cref{eq:securitystatement}, we first note that the protocol is  $\epsEV$-correct:
\begin{equation} \label{eq:protcorrec}
	\begin{aligned}
			\Pr(K_A \neq K_B \wedge \acceventEV) 		&\leq	\Pr(Z^n \neq \bobkey \wedge  \acceventEV ) \\
		&\leq \Pr(\acceventEV | Z^n \neq \bobkey ) \leq \epsilon_{\text{EV}},
	\end{aligned}
\end{equation}
where $\bobkey$ denotes Bob's guess for Alice's raw key, and the second inequality follows from the fact that $K_A\neq K_B \implies Z^n \neq \bobkey$, while the final inequality follows from that the fact that error-verification step compares hashes of length $ \ceil {\log(1/\epsEV)}$.  

Furthermore, we obtain the following chain of inequalities using some technical lemmas from  \cite{dupuisEntropy2020,dupuisPrivacy2022,tomamichelQuantum2016} which we restate in \cref{appendix:technical}. The derivation of these inequalities is explained below. We obtain,
\begin{widetext}
\begin{equation} \label{eq:iidLHLfixed}
			\begin{aligned}
		& \frac{1}{2}	\Pr( \fullaccevent{i} ) \tracenorm{ \finalstate{i}  - \finalstateideal{i}} \\
		& \leq \frac{1}{2} \Pr(\fullaccevent{i} ) 		2^{ -\left(  \frac{\alpha-1}{\alpha}\right)   \left( H_\alpha (Z^n | C^n \Cat^m C_{E} C_V E^N)_{\rho| \widetilde{\Omega}_i  \wedge \Omega_{\text{EV}}} - l_i\right) + \frac{2}{\alpha}  - 1     } \\
			&\leq 	 \frac{1}{2} \Pr(\fullaccevent{i}) 2^{ -\left(  \frac{\alpha-1}{\alpha}\right)   \left( H_\alpha (Z^n | C^n E^n)_{\rho| \widetilde{\Omega}_i  \wedge \Omega_{\text{EV}}} - \leak{i} -\EVCost - l_i\right) + \frac{2}{\alpha}  - 1     }  		\\
			&\leq  \frac{1}{2}2^{ -\left(  \frac{\alpha-1}{\alpha}\right)   \left( H_\alpha (Z^n | C^n E^n)_{\rho} - \leak{i} -\EVCost - l_i\right) + \frac{2}{\alpha}  - 1     }  \\
				&=
			\frac{1}{2}2^{ -\left(  \frac{\alpha-1}{\alpha}\right)   \left( n H_\alpha (Z | C E)_{\rho} - \leak{i} -\EVCost - l_i\right) + \frac{2}{\alpha}  - 1     }  \\
			&\leq 
			\frac{1}{2}2^{ -\left(  \frac{\alpha-1}{\alpha}\right)   \left( n H (Z | C E)_{\rho} -n (\alpha-1) \log^2(d_Z+1) - \leak{i} - \EVCost - l_i\right) + \frac{2}{\alpha}  - 1     }  \\
			& \leq \epsPA \quad \forall \rho \in \widetilde{S}_i,
			\end{aligned}
		\end{equation}
		\end{widetext}
		where $H_\alpha$ denotes the \renyi  entropy (see \cref{def:renyientropy,appendix:technical}) with $\alpha$ as the \renyi parameter. Here we used the leftover hashing lemma for \renyi entropy  \cite[Theorem 8]{dupuisPrivacy2022} (restated in  \cref{lemma:LHLrenyi}) in the first inequality, and \cref{lemma:renyisplitting} to split off the error-correction and error-verification information for the second inequality, along with the registers $\Cat^m E^{m}$ (which are independent of $Z^n$). We further use \cref{lemma:renyiconditioning} to get rid of the conditioning on acceptance events in the third inequality. The fourth equality follows from additivity of \renyi entropy  (\cref{lemma:additivity}), and  fifth inequality follows from  \cref{lemma:contrenyi} . The choice of $l_i$ from \cref{eq:l_ivalue} is the largest possible value that guarantees the final inequality in \cref{eq:iidLHLfixed}. The IID assumption comes into play in the use of \cref{lemma:additivity}.

Since either $\rho \in \Sset{i}$ or $\rho \notin \Sset{i}$, \cref{eq:Sdef,eq:iidLHLfixed} together imply that the protocol is $	\max\{ \epsAT, \epsPA\}$-secret:	\begin{equation} \label{eq:protsecrecy}
 	\begin{aligned}
		& \frac{1}{2}	\Pr(\fullaccevent{i} ) \tracenorm{ \finalstate{i} - \finalstateideal{i} } \\
		& \leq  \max\{ \epsAT, \epsPA\},
		\end{aligned}
	\end{equation}
	for all states $\Tr_{B^NE^N}(\rho^{\otimes N}_{ABE} ) = \bar{\sigma}^{\otimes N}_A$.
 Finally, as previously mentioned, a fixed-length QKD protocol that is $\epsEV$-correct (\cref{eq:protcorrec}) and $\max\{ \epsAT, \epsPA\}$-secret, is  $(\max\{ \epsAT, \epsPA\}+\epsEV)$-secure (\cref{eq:securitystatement}), as desired.

	\subsection{From fixed-length security to variable-length security}
	\label{subsec:fromfixedtovariable}
	
	For a variable-length protocol, again following composable security definitions~\cite{ben-orUniversal2004,portmannSecurity2022}, we say that it is $\epsSec$-secure against some class of attacks~\footnote{A full specification of a composable security framework~\cite{portmannCryptographic2014a,portmannSecurity2022} also technically requires describing some \emph{honest} ideal functionality in the case where Eve does not attack the protocol. For a variable-length protocol, we can take this to simply be a functionality that outputs perfect keys (of variable length) to Alice and Bob and nothing to Eve except the length of the key, with the distribution of key lengths being the same as that of the honest protocol behaviour. (This behaviour does not have to be explicitly known, for instance when considering the \cref{sec:truevariablelength} protocol. We merely require this honest behaviour to \emph{exist} in principle, and (to avoid only having trivial operational implications) for it to produce some ``reasonable'' expected key rate.) With this choice of ideal functionality, the protocol satisfies the property of \textit{completeness} (see~\cite{portmannCryptographic2014a,portmannSecurity2022} for details) with perfect completeness parameter.} if the following condition holds: for all attacks in that class, at the end of the protocol we have
	\begin{equation} \label{eq:secdef}
		\begin{aligned}
		&\sum_{k=0}^\infty \frac{1}{2}	\Pr(\Omega_{\mathrm{len}=k} ) \tracenorm{ \rho^{(k)}_{K_A K_B \Cfull E^N| \Omega_{\mathrm{len}=k} } - \rho^{(k,\text{ideal})}_{K_A K_B \Cfull E^N| \Omega_{\mathrm{len}=k} } } \\
		 &\leq \epsSec.
		\end{aligned}
	\end{equation}
	Here, $\Omega_{\mathrm{len}=k}$ denotes the event that a final key~\footnote{In principle there is the technicality that for an arbitrary variable-length protocol, Alice and Bob might produce final keys of different lengths. For this work, we focus on protocols where the final key length is completely determined from the public announcements, and so this is not an issue.} of length $k$ is produced, while $\rho^{(k)}_{K_A K_B \Cfull E^N| \Omega_{\mathrm{len}=k} } $ denotes the actual output state at the end of the protocol conditioned on the event $\Omega_{\mathrm{len}=k}$, and $\rho^{(k,\text{ideal})}_{K_A K_B \Cfull E^N| \Omega_{\mathrm{len}=k} } $ denotes an ``ideal state'' obtained by replacing the key registers of $\rho^{(k)}_{K_A K_B \Cfull E^N| \Omega_{\mathrm{len}=k} } $ with perfect keys of length $k$ (analogous to \cref{eq:idealfixedlength}). Note that we recover the security definition of fixed-length protocols (\cref{eq:securitystatement}) from \cref{eq:secdef} by setting $k$ to be a fixed value $k=l_i$ in the sum in \cref{eq:secdef}, and noting that $\Omega_{\mathrm{len} = k}$ is the same event as $ \fullaccevent{i}$, corresponding to a key length of $l_i$ bits, and $\leak{i}$ bits used for error-correction. 
	Again, in this work we focus only on IID collective attacks, in the sense previously described in \cref{subsec:fixedlengthstatements}.
	
	Similar to fixed-length protocols, one can define correctness and secrecy for variable-length protocols. Specifically, we shall take $\epscorrect$-correctness to be defined the same way as before (\cref{eq:corrdef}), while $\epssecret$-secrecy is analogously defined by omitting Bob's registers from the variable-length $\epsSec$-security condition, i.e.~for all attacks (in the considered class) we have
	\begin{equation} \label{eq:secretdefvar}
		\begin{aligned}
		&\sum_{k=0}^\infty \frac{1}{2}	\Pr(\Omega_{\mathrm{len}=k} ) \tracenorm{ \rho^{(k)}_{K_A \Cfull E^N| \Omega_{\mathrm{len}=k} } - \rho^{(k,\text{ideal})}_{K_A \Cfull E^N| \Omega_{\mathrm{len}=k} } } \\
	 &\leq \epssecret.
		\end{aligned}
	\end{equation}
	Just as in the fixed-length case, for the variable-length case we also have the property that $\epscorrect$-correctness and $\epssecret$-secrecy together imply $(\epscorrect + \epssecret)$-security --- the argument is identical to the fixed-length case \cite{portmannCryptographic2014a,portmannSecurity2022,ben-orUniversal2004}, and we provide it in \cref{lemma:correctandsecret}  of Appendix.~\ref{appendix:technical}.

	In order to use the security statements from \cref{subsec:fixedlengthstatements} for (a sequence of) fixed-length protocols to prove security for a variable-length protocol, we require the acceptance sets $\accset{i}$ of those fixed-length protocols to satisfy the following condition. We assume that the acceptance sets $\accset{i}$ for the fixed-length protocols are ordered such that $\accset{i} \subseteq \accset{i+1}$. This can in principle be satisfied by suitable construction of the acceptance sets, as we show in \cref{sec:applicationbb84} (though the resulting variable-length protocol may not be suitable in all contexts, as we discuss later in~\cref{sec:truevariablelength}).   Without loss of generality, we can then pick feasible sets $\Sset{i}$ such that $\Sset{i} \subseteq \Sset{i+1}$,  since 
		 \begin{equation} 
		 	\rho \notin \Sset{i+1} \implies \Pr(\accevent{i+1}) \leq \epsAT \implies \Pr(\accevent{i})  \leq \epsAT .	
		 \end{equation}Thus, the feasible set $\Sset{i}$ can always be chosen to be smaller than the feasible set $\Sset{i+1}$.

\begin{remark} \label{remark:decreasing} Recall from \cref{eq:l_ivalue}, we have $l_i  = \floor{ n \min_{\rho \in \widetilde{S}_i} H(X | C E)_{\rho} - \leak{i}- \text{constant correction terms} } $. Thus, $\Sset{i} \subseteq \Sset{i+1}$ implies that $l_i + \leak{i}$ is a \textit{non-increasing sequence in $i$}.  This property will play a crucial part in proving the security of our variable-length protocol. \end{remark}

 We now use the acceptance sets $\accset{i}$ from the sequence of fixed-length protocols to construct the sets $Q_i$ (in the variable-length decision step) for a variable-length protocol. Specifically, let us define $Q_1 \coloneqq \accset{1}$, and $ Q_i \coloneqq \accset{i} \setminus \accset{i-1}$.  
We then prove the following theorem concerning the security of variable-length QKD protocols.

	\begin{theorem} \label{thm:main}
	Let there be a sequence of fixed-length QKD protocols that vary only in their acceptance criterion ($\accset{i}$), length of error-correction communication ($\leak{i}$)  and final hash length $l_i$. Suppose that for each of these fixed-length protocols, we have a security proof against IID collective attacks, in which \cref{eq:Sdef,eq:l_ivalue,eq:protcorrec} are true.	Furthermore, suppose $\accset{i}\subseteq \accset{i+1}$ and $\Sset{i} \subseteq \Sset{i+1}$.	Then, the variable-length protocol that upon the event $\fullacceventadapt{i}$, generates a key of length $l_i$ while having used $\leak{i}$ number of bits for error-correction, is  $(\epsAT+\epsEV + \epsPA)$-secure against IID collective attacks, where the values of $\epsAT,\epsEV,\epsPA$ are the same as those in the fixed-length protocol statements \cref{eq:Sdef,eq:l_ivalue,eq:protcorrec}. 
	
	\end{theorem}
	
	\begin{proof}

	 As before, we will prove that the protocol is $\epsEV$-correct and $(\epsAT+\epsPA)$-secret. This will then imply that the protocol is  $(\epsEV+\epsAT+\epsPA)$-secure.
	 
	  The proof of $\epsEV$-correctness of the protocol remains essentially the same as before (\cref{eq:protcorrec}):

	 	\begin{equation} \label{eq:adaptcorrect}
	 		\begin{aligned} 
	 			\Pr(K_A\neq K_B \wedge \acceventEV ) &\leq \Pr(Z^n \neq \bobkey \wedge \acceventEV ) \\ 
	 			& \leq \Pr(\acceventEV | Z^n \neq \bobkey) \\
	 			& \leq \epsEV.
	 		\end{aligned}
	 	\end{equation}
	 	
	 	 We now focus on proving the $(\epsAT+\epsPA)$-secrecy of the protocol. To do so, we first
	 	 note that the secrecy definition for variable-length protocols (\cref{eq:secretdefvar}) groups together terms with the same output length of the key. However, the different events $\fullacceventadapt{i}$ may correspond to the same output length of the key and different lengths of error-correction information. Nevertheless,  the events $\fullacceventadapt{i}$ are deterministic functions of public announcements $\Cat^m,C_V$. Thus, the states $\finalstateadapt{i}$ conditioned on event $\fullacceventadapt{i}$ have orthogonal supports. Therefore, it is sufficient to show that
	 	 	 	\begin{equation} \label{eq:secretdefvar2}
	 	 	 		\begin{aligned}
	 	 	 		&\sum_{i=1}^M  \half \Pr(\fullacceventadapt{i} ) \tracenorm{ \finalstateadapt{i}- \finalstateidealadapt{i}} \\
	 	 	 		& \leq \epsAT + \epsPA.
	 	 	 		\end{aligned}
	 	 	 	\end{equation}
	 	  since we can group terms with the same output key length in \cref{eq:secretdefvar2} to show that  \cref{eq:secretdefvar,eq:secretdefvar2} are equivalent. (An analogous argument can be conducted at the level of the security condition (\cref{eq:secdef}) directly, though here we focus on just the secrecy condition since that (together with correctness) is sufficient to imply the security condition.) We will now prove \cref{eq:secretdefvar2}.

	 	 We proceed by noting that since we have the ordering $\Sset{i} \subseteq \Sset{i+1}$, any input state $\rho_{AB}$ has to fall under exactly one of the following three cases:
	 	\begin{enumerate}
	 		\item  $\rho_{AB} \in \widetilde{S}_1$. 
	 		\item $\rho_{AB} \notin \widetilde{S}_j$ but $\rho \in \widetilde{S}_{j+1}$ for some $j$.
	 		\item $\rho_{AB} \notin \widetilde{S}_M$.
	 			\end{enumerate}
	 			We prove the secrecy claim separately for each case. We start with Case 2.			

 \textbf{Case 2:}  If $\rho \notin \widetilde{S}_j$ but $\rho \in \widetilde{S}_{j+1}$ for some $j$, we split up the security definition from \cref{eq:secdef} into two parts. For the first part, we show that if $\rho \notin \Sset{j}$, the probability of the protocol obtaining the event $\Omega_1 \cup ... \Omega_{j}$ is small:
	\begin{equation} \label{eq:part1}
	\begin{aligned}
		&\sum_{i=1}^j\half	\Pr(\fullacceventadapt{i} ) \tracenorm{ \finalstateadapt{i}- \finalstateidealadapt{i}}  \\
		   &\leq 	\sum_{i=1}^j 	\Pr(\fullacceventadapt{i} ) \leq 	\sum_{i=1}^j	\Pr(\Omega_i  ) = \Pr(\accevent{j}) \leq \epsAT,
	\end{aligned}
\end{equation}
	where the final inequality follows from the fixed-length security statement  (\cref{eq:Sdef}). \\\\
 To bound the remaining terms, we use some technical lemmas from  \cite{dupuisEntropy2020,dupuisPrivacy2022,tomamichelQuantum2016} (which are restated in \cref{appendix:technical}), to obtain the following chain of inequalities.
 	
 	\begin{widetext}
	\begin{equation} \label{eq:part2}
		\begin{aligned}
			&\sum_{i=j+1}^M \half \Pr(\fullacceventadapt{i} ) \tracenorm{ \finalstateadapt{i}- \finalstateidealadapt{i}}    \\
			 &\leq 	\sum_{i=j+1}^M 	\frac{1}{2} \Pr(\fullacceventadapt{i} )2^{ -\left(  \frac{\alpha-1}{\alpha}\right)   \left( H_\alpha (Z^n | C^n \Cat^m C_{E}C_V E^N)_{\rho| \fullacceventadapt{i}} - l_i\right) + \frac{2}{\alpha}  - 1     }  \\
			  &\leq 	\sum_{i=j+1}^M 	\frac{1}{2} \Pr(\fullacceventadapt{i})2^{ -\left(  \frac{\alpha-1}{\alpha}\right)   \left( H_\alpha (Z^n |  C^n \Cat^m  C_V E^n)_{\rho| \Omega_i  \wedge \Omega_{\text{EV}}} - \leak{i} -  l_i\right) + \frac{2}{\alpha}  - 1     }  \\
			  &\leq 	\sum_{i=j+1}^{M} 	\frac{1}{2} \Pr(\fullacceventadapt{i})2^{ -\left(  \frac{\alpha-1}{\alpha}\right)   \left( H_\alpha (Z^n | C^n \Cat^m C_V  E^n)_{\rho| \fullacceventadapt{i}} - \leak{j+1} -  l_{j+1}  \right) + \frac{2}{\alpha}  - 1     }  \\
			  &\leq 		\frac{1}{2} 2^{ -\left(  \frac{\alpha-1}{\alpha}\right)   \left( H_\alpha (Z^n | C^n \Cat^m C_V  E^n)_{\rho} - \leak{j+1} -  l_{j+1}  \right) + \frac{2}{\alpha}  - 1     }  \\
			  & \leq 	\frac{1}{2} 2^{ -\left(  \frac{\alpha-1}{\alpha}\right)   \left( H_\alpha (Z^n | C^n   E^n)_{\rho} - \leak{j+1} - \EVCost -  l_{j+1}  \right) + \frac{2}{\alpha}  - 1     }  \\
			 & =	\frac{1}{2} 2^{ -\left(  \frac{\alpha-1}{\alpha}\right)   \left(  n H_\alpha (Z | C  E)_{\rho} - \leak{j+1} - \EVCost -  l_{j+1}  \right) + \frac{2}{\alpha}  - 1     } \\
			 & \leq 	\frac{1}{2} 2^{ -\left(  \frac{\alpha-1}{\alpha}\right)   \left( n  H(Z | C   E)_{\rho} - n(\alpha-1) \log^2(d_Z+1) - \leak{j+1} - \EVCost -  l_{j+1}  \right) + \frac{2}{\alpha}  - 1     } \\
			  &\leq \epsPA.
		\end{aligned}
	\end{equation}
	\end{widetext}
	The inequalities above are explained below, with the crucial step explained in \cref{remark:technicalreason}. 
	We used the leftover hashing lemma for \renyi entropy in the first inequality (\cref{lemma:LHLrenyi}), and \cref{lemma:renyisplitting} to split off the information leakage due to error-correction and the $E^{m}$ register (which is independent of $Z^n$), in the second inequality. For the third inequality, we use the fact that $l_i + \leak{i}$ is a non-increasing sequence in $i$ (\cref{remark:decreasing}).  We use \cref{lemma:renyiweightedaverage} to get rid of the conditioning on events for the fourth inequality, and \cref{lemma:renyisplitting} to split off information leakage due to error-verification and the $\Cat^m$ register (which is independent of $Z^n$) in the fifth inequality. The sixth equality follows from the additivity of \renyi entropy (\cref{lemma:additivity}), while the seventh inequality follows from \cref{lemma:contrenyi}. Finally, we use the security proof statement for fixed-length protocols (\cref{eq:iidLHLfixed}) and the fact that $\rho \in \widetilde{S}_{j+1}$ for the final inequality. 
	
	\begin{remark} \label{remark:technicalreason}
		We highlight two critical steps in \cref{eq:part2}. The first is in the third inequality, where we replace $l_i+\leak{i}$ with the constant value $l_{j+1}+\leak{j+1}$, using \cref{remark:decreasing}. The second is in the use of \cref{lemma:renyiweightedaverage} in the fourth inequality, which allows us to get rid of terms involving \renyi entropies of the state conditioned on events. In particular, smooth min-entropy does not straightforwardly allow a statement analogous to \cref{lemma:renyiweightedaverage}, which is the reason for using \renyi entropy in this work.
		Moreover  we split off the $\Cat^m,C_V$ registers \text{after} using \cref{lemma:renyiweightedaverage}, since we require the events $\fullacceventadapt{i}$ to be known to Eve to use \cref{lemma:renyiweightedaverage} for the fourth inequality. 	\end{remark}
	Bringing Eqs.~\eqref{eq:part1} and~\eqref{eq:part2} together, we obtain that the protocol is $(\epsAT+\epsPA)$-secret:
		\begin{equation}
			\begin{aligned}
	&	\sum_{i=1}^M  \frac{1}{2}	\Pr(\Omega_i \wedge \Omega_{\text{EV}} ) \tracenorm{  \finalstateadapt{i} - \finalstateidealadapt{i}} \\
		&\leq  \epsilon_{\text{AT}}+ \epsilon_{\text{PA}} .
		\end{aligned}
	\end{equation}
	
	\textbf{Case 1 and Case 3:} The analysis of Case 1 is a special case of the above analysis, and follows from choosing $j=0$ in \cref{eq:part2}. The analysis of Case 3 is also a special case, and follows from choosing $j=M$ in \cref{eq:part1}.
	
	The theorem claim then follows from the correctness and secrecy statements.
	\end{proof} 
	Thus, a sequence of security proofs for fixed-length protocols satisfying certain conditions can be turned into a security proof for a variable-length protocol. Moreover, the only penalty imposed by our approach is a minor increase in the security parameter of the protocol, which goes from $(\max \{\epsAT,\epsPA\}+\epsEV)$ for the fixed-length case, to $(\epsAT+\epsPA+\epsEV)$ for the variable-length case. 
	In fact, this minor penalty is completely compensated by the ability to generate longer keys in the variable-length case, as we show in the next section.

	\section{Application to Qubit BB84} \label{sec:applicationbb84}
	In this section we will show how  \cref{thm:main} can be utilized to improve the \textit{expected key rate} \cite{kanitscharFiniteSize2023a} of QKD protocols.  For the sake of simplicity, we consider the qubit based BB84 protocol to illustrate our results. However, \cref{thm:main} can be directly applied to any fixed-length protocols whose security proof satisfies \cref{eq:Sdef,eq:l_ivalue,eq:protcorrec}.
	We use Ref.~\cite{georgeNumerical2021} for the finite-size security proof of qubit BB84, and the numerical key rate framework from Ref.~\cite{winickReliable2018} to compute key rates. 
The signal preparation and measurement steps of the qubit BB84 protocol are described in \cref{appendix:bb84}. The acceptance test and key rate computation is described in \cref{subsec:ianacceptancetest}. 
We start by explaining the notion of expected key rates.

 \subsection{Expected key rate}
  Before defining expected key rates we first set up the following notation. 
 
 \begin{enumerate}
 	\item  $\KRfixed{i}$: This denotes the key rate obtained upon the event $\fullaccevent{i}$ for the $i$th fixed-length protocol.
 	\item $\KRvariable{i}$: This denotes the  key rate obtained upon the event $\fullacceventadapt{i}$ for the variable-length protocol.
 	\item $\bar{R}_\text{fixed,i}$: This denotes the  expected key rate for the $i$th fixed-length protocol.
 	\item $\bar{R}_\text{variable}$: This denotes expected key rate for the variable-length protocol.
 	\item $\honest$: This denotes the state corresponding to the honest implementation of the QKD protocol. 
 \end{enumerate}
Note that $\KRfixed{i}$ and $\KRvariable{i}$ are obtained from the security proofs, and are independent of honest behaviour. 

For the purposes of this work, in all the expected key rate computations we assume that the probability of the event $\acceventEV$ in the honest case is approximately $1$. We make this simplifying assumption because the true value would depend on the (honest) probability of Bob correctly guessing Alice's key in the error-correction step, but many error-correction protocols used in practice do not have rigorous lower bounds on this probability, only heuristic estimates. We stress however that this in no way affects our proof that the protocol satisfies the \emph{security} condition, which does not require any lower bound on this probability.

With this approximation, the expected key rate for the $i$th fixed-length protocol is given by
\begin{equation} \label{eq:expkrfixed}
	\begin{aligned}
		\bar{R}_\text{fixed,i} &\coloneqq  \Pr(\accevent{i} \wedge \acceventEV )_{\honest} \KRfixed{i} \\
		& \approx \Pr(\accevent{i})_{\honest} \KRfixed{i},
	\end{aligned}
\end{equation}
where $\Pr(\accevent{i})_{\honest}$ is the probability of the protocol accepting during honest behaviour, and $\KRfixed{i}$ is the key rate upon accepting for the  $i$th protocol. 
We use $\bar{R}_\text{fixed}$ as a useful metric to compare the \textit{practical key rate} of a QKD protocol. 
 
We can generalize the notion of expected key rate to the variable-length case in a straightforward manner.  We define 
\begin{equation} \label{eq:expkrvariable}
	\begin{aligned}
		\bar{R}_\text{variable}  &\coloneqq \sum_{i=1}^M \Pr(\fullacceventadapt{i})_{\honest} \KRvariable{i} \\
		&\approx \sum_{i=1}^M \Pr(\acceventadapt{i})_{\honest} \KRvariable{i},
		\end{aligned}
\end{equation}
where $\Pr(\acceventadapt{i})_{\honest}$ is the probability of obtaining the event $\acceventadapt{i}$ for honest implementations, and $\KRvariable{i}$ is the key rate obtained upon the event $\acceventadapt{i}$. Again, $\bar{R}_\text{variable}$ is a useful metric to compare the \textit{practical key rate} of a QKD protocol. 

Thus, the expected key rates can be computed from \cref{eq:expkrfixed,eq:expkrvariable}. The values of $R_\text{fixed,i}$ and $R_\text{variable,i}$ can be obtained from security proofs. The probabilities $ \Pr(\accevent{i})_{\honest}$ and $ \Pr(\acceventadapt{i})_{\honest}$ can be estimated numerically, by simulating the channel a large number of times, and computing the fraction of runs that lead to events $\accevent{i}$ and  $\acceventadapt{i}$.

We now describe the acceptance test from  Ref.~\cite{georgeNumerical2021}, and key rate computations.

		\subsection{Acceptance Test and Key Rates} \label{subsec:ianacceptancetest} 
				Consider a sequence of fixed-length protocols indexed by $i \in \{1,2,\dots,M\}$. Let $\Sigma $ denote the set of outcomes that can take place in the test rounds. For qubit BB84,  $\Sigma$ consists of the $16$ possible outcomes corresponding to Alice's choice of signal state and Bob's measurement outcome. Then following Ref.~\cite{georgeNumerical2021}, we shall define the acceptance set $\accset{i}$ for each of these fixed-length protocols as
				\begin{equation} \label{eq:acceptancetest}
					\begin{aligned}
						\widetilde{Q}_i \coloneqq	 \{ \Fobs \in \mathcal{P}(\Sigma) \; | \; \norm{\Fobs - \Fbar }_1 \leq t_{i}  \} ,
					\end{aligned}
				\end{equation}
				where $\mathcal{P}(\Sigma)$  the set of probability distributions on $\Sigma$. Here $\Fbar$ is the probability vector of outcomes for the honest implementation, i.e. each entry of $\Fbar$ is the probability of obtaining some outcome in a single round of the honest implementation, as determined by \cref{eq:Phimap} below. 
				$\Fobs$ is the observed frequency of outcomes, and the acceptance test checks whether the observed frequency of outcomes is close to the expected frequency ($\Fbar$). 			Note that one can easily satisfy the condition $\accset{i} \subseteq \accset{i+1}$, by choosing $t_{i} \leq t_{i+1}$. This ensures that \cref{thm:main} can be applied safely. 
				
				Let $\Gamma_j$ be the POVM element corresponding to the $j$th outcome, and define 
				\begin{equation} \label{eq:Phimap}
					\Phi(\rho) \coloneqq \sum_{j \in \Sigma} \Tr (\Gamma_j \rho) \ket{j} \bra{j}
				\end{equation}
				to be the map that takes the state $\rho$ and outputs the  probability distribution over the outcomes. 				
				Given such an acceptance set $\accset{i}$, a feasible set $\Sset{i}$ satisfying \cref{eq:Sdef} is given by \cite[Theorem 8]{georgeNumerical2021}
				\begin{equation} \label{eq:bb84Sset}
					\Sset{i} \coloneqq \{ \rho \in S_\circ(AB) |  \norm{\Phi(\rho) - \Fbar }_1 \leq t_i + \mu \},
				\end{equation}
				where $\mu$ is given by
				\begin{equation}
						\mu \coloneqq \sqrt{2} \sqrt{ \frac{\ln(1/\epsAT)+|\Sigma| \ln(m+1)} {m}}.
				\end{equation}
				The construction of this set crucially uses the concentration inequality from \cref{lemma:concentration} (\cref{appendixsubsec:acceptancetest}). 
				
				Therefore, the key length $l_i$ satisfying \cref{eq:l_ivalue} is given by		 
				\begin{equation} \label{eq:bb84keyrate}
				\begin{aligned}
					l_i \leq \; & n  \min_{ \substack{\rho \in \Sset{i} \\ \Tr_B(\rho) = \bar{\sigma}_A}  }H(Z|CE)_{\rho}  - \leak{i} -  n (\alpha-1) \log^2(2d_Z+1) \\
					&-  \EVCost -  \frac{\alpha}{\alpha-1} \left( \log \left( \frac{1}{4\varepsilon_{\text{PA}}}\right) +  \frac{2}{\alpha} \right) ,
					\end{aligned}
				\end{equation}
				where $d_Z$ is the dimension of $Z$, $\leak{i}$ is the number of bits used for error-correction, $n$ is the number of signals for key generation, and we set the \renyi parameter to be $\alpha = 1 + \kappa/\sqrt{n}$ with $\kappa \coloneqq \sqrt{{\log(1/\epsilon_\text{PA})}}/{\log(d_Z+1)}$.  Moreover, $\leak{i}$ can be chosen to be any number for the purposes of proving the security of the protocol. However, a careful choice of $\leak{i}$ and design of the error-correction protocol is necessary to guarantee that the protocol passes error-verification with high probability for honest behaviour.

		\subsection{Results}
		\label{subsec:bb84resultsadaptive}
				\begin{figure}
			\includegraphics[width=\linewidth]{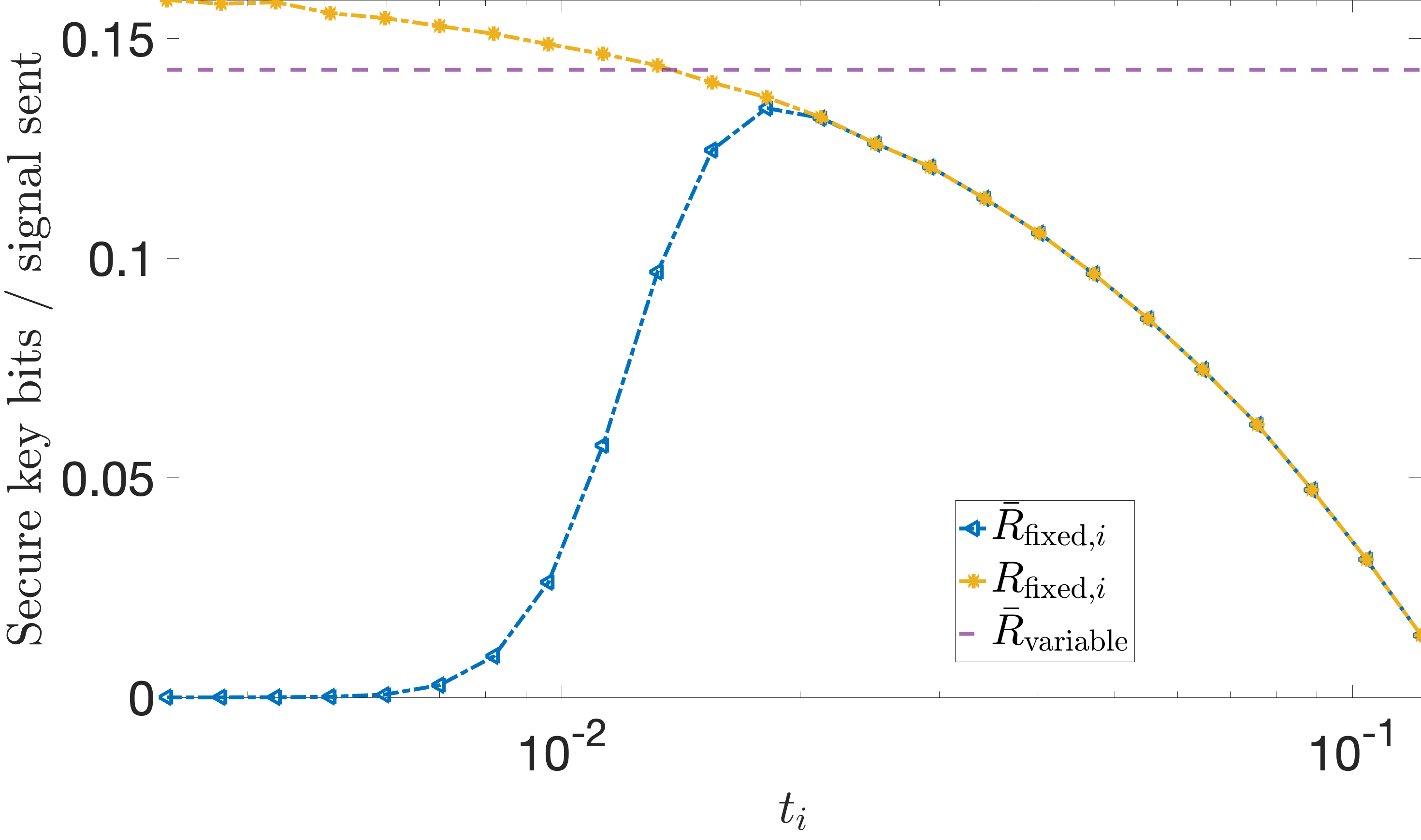} 
			\caption{Expected key rate for fixed-length protocols ($\bar{R}_{\text{fixed},i}$) for various values of $t_i$, key rate upon acceptance for fixed-length protocols ($\KRfixed{i}$) for various values of $t_i$, and the expected key rate for the variable-length protocol ($\bar{R}_{\text{variable}}$) constructed from the fixed-length protocols. } \label{fig:expkeyrateplot} 
		\end{figure}
		
		We now use the above results to compare the expected key rates for fixed-length protocols and variable-length protocols. 
		We consider a protocol with honest behaviour determined by a depolarization probability of $0.02$, and misalignment angle about the $Y$ axis of $\theta=2^\circ$. We set the basis choice probabilities 
		to $p_z=p_x=0.5$. The total number of signals is given by $N=10^6$, and the number of signals used for testing is given by $m=0.05N$. The number of bits used for error-correction is always taken to be $\leak{i} = \leak{} =  f n H(Z | YC)_{\honest}$ where $f=1.16$ is the efficiency parameter. We set {$\alpha = 1 + \kappa/\sqrt{n}$ with $\kappa \coloneqq \sqrt{{\log(1/\epsilon_\text{PA})}}/{\log(d_Z+1)}$, and fix a range of values of $t_i$ (horizontal axis of \cref{fig:expkeyrateplot}) such that $t_i \leq t_{i+1}$. This determines the acceptance sets $\accset{i}$ for the fixed-length protocols, and thus also the sets $\accsetadapt{i}$ in the variable-length decision of the variable-length protocol we construct (\cref{sec:fixedtoadaptive}).  We set a target security parameter of $\epsSec = 10^{-12}$.  We plot various key rates in \cref{fig:expkeyrateplot}, which are explained below.
		
		\begin{enumerate}
			\item $\KRfixed{i}$: This is the key rate upon acceptance for fixed-length protocols for various values of $t_i$.  Since $\epsSec = \max\{\epsPA,\epsAT\}+\epsEV$
			for fixed-length protocols, we set $\epsAT =\epsPA = \epsEV = \epsSec/2$. We use \cref{eq:bb84keyrate,eq:bb84Sset} to compute $l_i$ for various values of $t_i$, and plot  $\KRfixed{i} = l_i / N$. We see that $\KRfixed{i}$ decreases monotonically on increasing values of $t_i$, which reflects the fact that larger acceptance sets lead to lower key rate upon acceptance. 
			\item $\bar{R}_{\text{fixed},i}$: This is the expected key rate for fixed-length protocols, for various values of $t_i$.  We use the values of $\KRfixed{i}$ obtained above, along with \cref{eq:expkrfixed} to compute $\bar{R}_{\text{fixed},i}$.  The probability of accepting the protocol $\Pr(\accevent{i})_{\honest}$ is computed as follows:
			\begin{enumerate}
				\item We first compute the probability vector $\Fbar$ corresponding to the honest behaviour $\honest$, by setting $\Fbar=\Phi(\honest)$, where $\Phi$ is defined in  \cref{eq:Phimap}.
				\item We sample $m$ times from  $\Fbar$, and obtain the observed frequency of outcomes $\Fobs$. We check whether $\Fobs \in \accset{i}$.
				\item  We estimate $\Pr(\accevent{i})_{\honest}$ by repeating (b) $100000$ times, and computing the fraction of times we obtained $\Fobs \in \accset{i}$.
			\end{enumerate}
			
			We see that the $\bar{R}_\text{fixed,i}$ is small at low values of $t_i$, since the probability of the protocol accepting is small. We also see that $\bar{R}_\text{fixed,i}$ is small at larger values of $t_i$, since the key rate upon acceptance is small. The expected key rate thus captures the trade-off between accepting with large probability, versus producing a large key upon acceptance.
			\item $\bar{R}_\text{variable}$: This is the expected key rate for variable-length protocols. Note that this is a fixed value and \textit{not} plotted as a function of $t_i$ (since we obtain a single variable-length protocol from a sequence of fixed-length protocols determined by the $t_i$s). Anticipating the use of \cref{thm:main}, we set $\epsAT =\epsSec/4$, $\epsPA = \epsSec/4$, and $\epsEV = \epsSec/2$. We compute $l_i$ using \cref{eq:bb84keyrate} for various values of $t_i$, and set $\KRvariable{i} = l_i/N$. Using \cref{thm:main}, we obtain that the  variable-length protocol constructed from the sequence of fixed-length protocols, for the given set of $t_i$s, is $\epsSec=(\epsAT+\epsPA+\epsEV)$-secure. We compute the various probabilities $\Pr(\acceventadapt{i})_{\honest}$ in the same manner as (2) above (by simulating $100000$ runs of the QKD protocol), and compute $\bar{R}_\text{variable}$ using \cref{eq:expkrvariable}.
		\end{enumerate}
		
		Crucially, we find that the variable-length protocol has higher expected key rate than the \textit{best fixed-length protocol}.  Since the variable-length protocol consists of exactly the same steps as the fixed-length protocol, and only differs in the parameters of the classical processing of the data, the implementation of the variable-length protocol does not impose \textit{any} additional difficulties, and is accompanied by an \textit{increase} in the expected key rate. In fact, we expect that implementing a variable-length protocol will almost always lead to an improvement in the expected key rate, as we argue in the following remark.

\begin{remark} \label{remark:improving}		
	 Consider any fixed-length protocol where the honest behaviour is given by $\honest$. Suppose that Alice and Bob choose to implement a fixed-length protocol, with parameters $l$, $\leak{}$ and $t$.
	 Now, consider the variable-length protocol for the same honest behaviour $\honest$, for the same choice of $\epsPA,\epsAT,\epsEV$, obtained by choosing $l_i$ according to $t_1 \leq t_2 ... \leq t$, and choosing $\leak{i} = \leak{}$.  Then since $l_1 \geq l_2 ... \geq  l$, using \cref{thm:main}, one is guaranteed to improve upon the expected key rate by switching to a variable-length protocol (albeit with a small increase in the security parameter). This is because the variable-length protocol always has some non-zero probability of producing keys of larger length when compared to the fixed-length protocol. We believe that in almost all cases, this improvement will remain even after choosing the same security parameter for both fixed-length and variable-length protocols (as we saw in \cref{fig:expkeyrateplot}).
	\end{remark}

	\section{A true variable-length protocol} \label{sec:truevariablelength}
		In the preceding section, we considered a scenario where the honest implementation of the protocol is fixed and known beforehand. However, in scenarios where the honest implementation varies unpredictably between each run of the protocol, it is not clear how Theorem \ref{thm:main} can be used to obtain good key rates. For instance, suppose that the channel has a 50\% chance of having honest behaviour $\honest$ (leading to statistics $\Fbar$) and $\honest^\prime$(leading to  statistics $\Fbar^\prime$), and  $\Fbar$ and $\Fbar^\prime$ are very different frequencies.  Then it is not clear how to choose suitable acceptance sets that: a) give good key rates and b) on which \cref{thm:main} can be applied. This is because the size of the acceptance test that includes both $\Fbar$ and $\Fbar^\prime$, is of the order of $\norm{ \Fbar -\Fbar^\prime }_1$, which can be quite large. This leads to low (or in many cases zero) key rate upon acceptance. Thus, we have not yet resolved the problem of unpredictable channels, which we shall now address in this section.  We note that one potential solution to this problem is to coarse-grain the acceptance data, and set the acceptance condition to be ``QBER is less than some fixed value". However, coarse-graining involves throwing away information, and has been shown to lead to suboptimal key rates \cite{wangNumerical2022}. 	Note that an unpredictable channel connecting Alice and Bob is already a significant issue for experimental implementations, which is sometimes incorrectly resolved by choosing the acceptance test ($\Fbar, t$) \textit{after} seeing the observed statistics $\Fobs$ (see \cref{rem:acceptancetest}).

In this section, we will propose and analyze a variable-length protocol that directly uses $\Fobs$, the observed frequency of outcomes, to determine the length of the secret key to be produced and the number of bits to be used for error-correction. Note that unlike \cref{sec:fixedtoadaptive}, we no longer need to go through a sequence of fixed-length protocols in this section. Instead, we will design the variable-length decision in a different manner that does not depend on a sequence of fixed-length acceptance tests. Crucially, this will involve the construction of a statistical estimator $\bstat(\Fobs)$, that with high probability is a lower bound on the \renyi entropy $ H_\alpha(Z^n | C^nE^n )_{\rho}$ of the state $\rho$ in the QKD protocol. That is, we will first construct a $\bstat$ such that for any state $\rho$, it is the case that
	\begin{equation}
		\Pr_{\Fobs} \left( \bstat(\Fobs ) \leq H_\alpha(Z^n | C^nE^n )_{\rho} \right) \geq 1-\epsAT.
	\end{equation}
	This estimator will then be used to determine the length of the output key.  We start by presenting some results that allow us to construct such an estimator in \cref{subsec:bstat}. In  \cref{subsec:protspec} we specify the variable-length protocol, and explain how the users use $\bstat(\Fobs)$ to decide the length of the  key and the number of bits to use for error-correction. Finally, in \cref{subsec:thmsecond} we prove the security of the variable-length protocol. 
	
	We highlight that the only place we use the IID collective attacks assumption is in \cref{subsec:bstat}, in the construction of $\bstat$. Therefore, if alternative methods could be found that construct $\bstat$ without this assumption, our proof framework would generalize to coherent attacks.

		\subsection{ Constructing the Estimator } \label{subsec:bstat}
		In a QKD protocol, we deal with a \textit{fixed yet unknown} $\rho_{AB}$. In particular $\rho_{AB}$ is a fixed state and \emph{not} a random variable. This $\rho_{AB}$ then gives rise to a random variable $\Fobs$. Given that Alice and Bob observe $\Fobs$, obtained by performing measurements on $\rho_{AB}^{\otimes m}$, we would like to construct a set of states $V(\Fobs)$, such that $V(\Fobs)$ contains $\rho_{AB}$  with high probability. 
		
	\begin{remark} \label{remark:faircomparison}
			In general, one can use a variety of concentration inequalities to obtain such a set. In the following Lemma, we will use the concentration inequality from \cref{lemma:concentration} (\cref{appendixsubsec:acceptancetest}). We make this choice since it is the same concentration inequality used in the construction of the feasible set \cref{eq:bb84Sset}, and we wish to make a fair comparison between the variable-length and fixed-length protocols. Thus, for our comparisons later in \cref{sec:resultstrueadaptive}, the acceptance tests for the fixed-length protocols and the variable-length decision in the variable-length protocol are designed using the \textit{same} concentration inequalities.
		\end{remark}
		
		\begin{lemma} \label{lemma:truevariable}
			For any state $\rho$, let $\Fobs \in \mathcal{P}(\Sigma)$ be the frequency vector obtained from measuring the state $m$ times, where $\Sigma$ is the set of possible  outcomes. Let $\Gamma_j$ be the POVM element corresponding to outcome $j$. Define parameters
			\begin{equation} \label{eq:confinterval}
				\begin{aligned}
					\mu \coloneqq \sqrt{2} \sqrt{ \frac{\ln(1/\epsAT)+|\Sigma| \ln(m+1)} {m}},
			\end{aligned}
			\end{equation}
			and the map $\Phi(\rho) \coloneqq \sum_{j \in \Sigma} \Tr(\Gamma_j \rho) \ket{j} \bra{j}$, and the set
			 \begin{equation} \label{eq:Vset}
					V(\Fobs) \coloneqq \{  \rho \in  S_\circ(AB) \; | \;   \norm{\Phi(\rho) - \Fobs}_1 \leq \mu \}.
			\end{equation}
			Then, $V(\Fobs)$ contains $\rho$ with probability greater than $1-\epsAT$. That is,
			\begin{equation}  \label{eq:VFobscondition}
				\Pr_{\Fobs} \left( \rho \in V(\Fobs) \right) \geq 1-\epsAT.
			\end{equation}
			\end{lemma}

			\begin{proof}
					$\Fobs$ is sampled from the probability distribution given by $\Phi(\rho)$.
				The claim follows from \cref{lemma:concentration}, which states that if $\Fobs$ is obtained by sampling $m$ times from the probability distribution $\Phi(\rho)$, then
				\begin{equation}
					\Pr ( \norm{\Fobs - \Phi(\rho) }_1 \leq \mu ) \geq 1-\epsAT
					\end{equation}
				\end{proof}

				Next, we use the above result to obtain a statistical estimator of a lower bound on the \renyi entropy of the state $\rho^{\otimes n}_{ZCE}$. 
				\begin{lemma} \label{lemma:bstat}
						For any state $\rho$ satisfying $\Tr_{BE}(\rho_{ABE}) = \bar{\sigma}_A$, let $\Fobs \in \mathcal{P}(\Sigma)$ be the frequency vector obtained from measuring the state $m$ times. Define					
						\begin{equation}
							\bstat(\Fobs) \coloneqq \min_{\substack{\rho \in V(\Fobs), \\ \Tr_{BE}(\rho) = \bar{\sigma}_A} } \! \! n H(Z | C E)_\rho - n (\alpha-1) \log^2(d_Z+1).
						\end{equation}
					where $d_Z=\dim(Z)$ and $1 < \alpha < 1+ 1/\log(2d_Z+1)$. Then, 
						\begin{equation}
							\Pr_{\Fobs} \left( \bstat(\Fobs ) \leq H_\alpha(Z^n | C^nE^n )_{\rho} \right) \geq 1-\epsAT.
						\end{equation}
				\end{lemma}
				\begin{proof}
				
					 From \cref{lemma:contrenyi,lemma:additivity}, we have that $H_\alpha(Z^n | C^nE^n )_{\rho} \geq n H(Z | C E)_\rho - n (\alpha-1) \log^2(d_Z+1)$. The claim then follows from  \cref{lemma:truevariable}.
				\end{proof}
			
		\subsection{Variable length decision} \label{subsec:protspec} 
	 We will use $\bstat(\Fobs)$ to construct the following variable-length decision procedure. Let $\mathcal{F}$ be the (possibly infinite~\footnote{For protocols compatible with the specific $\bstat$ construction we use in this work, $\mathcal{F}$ would have to be finite because the formula we use in~\cref{eq:confinterval} requires the outcome space $\Sigma$ to be finite in order to obtain nontrivial results. However, we cover a possibly infinite $\mathcal{F}$ in this part of our analysis to accommodate potential follow-up work; in particular, for continuous-variable QKD it should be possible to construct an appropriate estimator $\bstat$ (via a different concentration inequality) even if the outcome space is infinite.}) set of all possible observations $\Fobs$ in the variable-length decision step.  Let $\{0,1,\dots,l_{\text{max}}\}$ denote all the possible values of output key lengths in our protocol. Let $\{0,1,\dots,\leak{\text{max}}\}$ denote all the possible values of the number of bits used for error-correction in our protocol. Then, the variable-length decision is implemented as follows:
				  \begin{enumerate}
				 	\item From public announcements $\Cat^m$, Alice and Bob compute $\Fobs$ and $\bstat(\Fobs)$ .
				 	\item They compute $\leak{}(\Fobs)$, the number of bits to be used for error-correction information, where  $\leak{}(\cdot) : \mathcal{F} \rightarrow \{0,1,\dots,\leak{\text{max}}\}$ is some predetermined function. 
				 	\item They compute $l(\Fobs)$, the length of the final key to be produced, where  $l(\cdot) : \mathcal{F} \rightarrow \{0,1,\dots,l_{\text{max}}\}$ is a function defined as 
				 	 \begin{equation}  \label{eq:livalue}
				 		\begin{aligned}
				 			l(\Fobs)& \coloneqq\max \bigg( 0,  \\
				 			&\left \lfloor \bstat(\Fobs) - \leak{}(\Fobs)  -\PAcost \right \rfloor \bigg), \\
				 			\PAcost &\coloneqq  \frac{\alpha}{\alpha-1} \left( \log(\frac{1}{4\epsPA} ) +  \frac{2}{\alpha}  \right) +  \EVCost.
				 		\end{aligned}
				 	\end{equation}
				 \end{enumerate}
				Recall that $\Omega_i$ denotes the event that a key of length $l_i$ is produced using $\leak{i}$ bits for error-correction for some values $(l_i, \leak{i})$. In other words, the index $i$ determines the pair $(l_i, \leak{i})$ of values of the key length and length of error-correction information. The sets $\accsetadapt{i}$ for the variable-length decision in our protocol are thus formally defined by
				 \begin{equation} \label{eq:accsetadapt}
				 	\accsetadapt{i} = \{ \Fobs \in \mathcal{F} | l(\Fobs) = l_i, \leak{}(\Fobs) = \leak{i} \}
				 \end{equation}
				 Note that number of possible events $\Omega_i$ is always finite (unlike the set of possible observations $\mathcal{F}$), and we denote it by $M$.
				 The remaining steps of the variable-length protocol are identical to the ones described in \cref{sec:notation}.

				\begin{remark} \label{remark:orderingfullyapadative}
					Recall that in the proof of \cref{thm:main}, the fact that $l_i+\leak{i}$ was a non-increasing sequence in $i$ played a crucial role. This property is required for similar reasons in the proof of \cref{thm:second}. Thus, without loss of generality, we label the events $\Omega_i$ such that 					$l_i+\leak{i}$ forms a non-increasing sequence in $i$.\end{remark}
				Such an ordering allows us to prove the following Lemma, which we use in the next section in our security proof.

	\begin{lemma} \label{lemma:ordering}
	Let $\mathcal{T}_p = \{ i | l_i > 0\}$ be the set of values of $i$ that lead to non-trivial length of the key. Then,
	\begin{equation}
		\begin{aligned}
		\sum_{\substack{i=1 \\ i\in \mathcal{T}_p}}^j \Pr( \Omega_i) \leq \Pr(\bstat(\Fobs) \geq l_j + \leak{j} + \PAcost ),
			\end{aligned}
	\end{equation}
	 for any $j \in \{1,2,\dots,M\}$.
	 
	\end{lemma}
	\begin{proof}
	For any $i \in \mathcal{T}_p$, using \cref{eq:livalue} we have
	\begin{equation}\label{eq:temp}
 \Omega_i \implies	\bstat(\Fobs) \geq l_i + \leak{i} +\PAcost
 \end{equation}
	Therefore
		\begin{equation}
			\begin{aligned}
		\sum_{\substack{i=1 \\ i \in \mathcal{T}_p} }^{j} & \Pr(\Omega_i)  =    \Pr(\bigcup_{\substack{i=1 \\ i \in \mathcal{T}_p} }^{j} \Omega_i ) \\
		&\leq \Pr(\bigcup_{\substack{i=1 \\ i \in \mathcal{T}_p}}^j \left( \bstat(\Fobs) \geq l_i + \leak{i}  +\PAcost \right)) \\
		&\leq \Pr( \bstat(\Fobs) \geq l_j + \leak{j}  +\PAcost )
		\end{aligned}
	\end{equation}
	where we used the fact that $\Omega_i$ are disjoint events in the first equality, \cref{eq:temp} for the second inequality, and the ordering on $l_i + \leak{i}$ (\cref{remark:orderingfullyapadative}) in the final inequality.
		\end{proof}
 We now have all the tools necessary to prove the security of our variable-length protocol. Before presenting the security proof, we compare the fixed-length and variable-length implementations in the following remark.
 \begin{remark} \label{remark:sets}
 		Note that with the way we construct the acceptance tests and variable-length decision in this section, the following property holds. Focusing on the fixed-length implementation for some specific $i$, the key length whenever the protocol accepts is given by \cref{eq:bb84keyrate}, which is an optimization over the feasible set $\Sset{i}$ (\cref{eq:bb84Sset}) whose size is determined by $t_i+\mu$.  On the other hand, the variable-length implementation determines the  key length (\cref{eq:livalue}) by looking at the observed value $\Fobs$ and optimizing over the set $V(\Fobs)$ (\cref{eq:Vset}) whose size is determined by $\mu$. Now observe that whenever $\Fobs$ takes a value such that the fixed-length implementation would accept during the acceptance test (\cref{eq:acceptancetest}), $V(\Fobs)$ is smaller than $\Sset{i}$, and the only difference between \cref{eq:bb84keyrate} and \cref{eq:livalue} is in the optimization set. Therefore, it follows that the variable-length key rate is always higher when the same values of $\epsAT,\epsEV, \epsPA$ are used in the two cases (though as previously discussed, this results a minor increase in $\epsSec$).
 	\end{remark}

	\subsection{Security proof of variable-length protocol} \label{subsec:thmsecond}

	\begin{theorem} \label{thm:second}
		The variable-length protocol that, on obtaining $\Fobs$ during the variable-length decision and passing error-verification, hashes to length $l(\Fobs)$ using $\leak{}(\Fobs)$ bits for error-correction (according to \cref{eq:livalue}), is $(\epsAT+\epsEV+\epsPA)$-secure.	
\end{theorem}

\begin{proof}

	As in the proof of \cref{thm:main}, we will show that the protocol is $\epsEV$-correct and $(\epsPA+\epsAT)$-secret, implying that the protocol is $(\epsEV+\epsAT+\epsPA)$-secure (\cref{lemma:correctandsecret} or Ref.~\cite{ben-orUniversal2004}). First note that the proof of $\epsEV$-correctness of the protocol is the same as in the proof of \cref{thm:main} (\cref{eq:adaptcorrect}). Thus we only need to prove secrecy.

	Again, as in the proof of \cref{thm:main}, it is  sufficient to show that 
		\begin{equation} \label{eq:secretdefvar3}
		\begin{aligned}
			&\sum_{i=1}^M  \half	\Pr(\fullacceventadapt{i} ) \tracenorm{ \finalstateadapt{i}  \! \!- \finalstateidealadapt{i} } \\
			& \leq \epsAT + \epsPA,
		\end{aligned}
	\end{equation}
	since each of the states $ \finalstateadapt{i} $ have orthogonal supports. This is because the event $\fullacceventadapt{i}$ is a deterministic function of the registers $\Cat^m, C_V$.  Thus, \cref{eq:secretdefvar3,eq:secretdefvar} are equivalent.

		  Recall that the values $l_i+ \leak{i}$ are ordered such that they form a non-increasing sequence (\cref{remark:orderingfullyapadative}).  Thus, for \textit{any} $\rho_{AB}$ that the protocol can start with, the \renyi entropy $H_\alpha(Z^n | C^n E^n)_{\rho}$ has to fall under at least one of the following three cases: 
			\begin{enumerate}
				\item $H_\alpha(Z^n | C^n E^n)_{\rho} \geq l_1+\leak{1}+\PAcost$.
				\item $ l_j + \leak{j} + \PAcost \geq H_\alpha(Z^n | C^n E^n)_{\rho} \geq l_{j+1} + \leak{j+1} + \PAcost$ for some $j \in \{1,...,M-1\}$.
				\item $ l_{M} + \leak{M} + \PAcost\geq  H_\alpha(Z^n | C^n E^n)_{\rho}$.
			\end{enumerate} 
			We will prove the secrecy claim separately for each case. Suppose $\rho$ is such that it satisfies case 2, for some value  $j$.   In this case, the secrecy bound can be obtained similar to the proof of Theorem \ref{thm:main}, by splitting up the sum into two convenient parts. The first part groups the set of events that happen with low probability, and is given by,
			 	\begin{equation} \label{eq:newpart1}
			 	\begin{aligned}
			 	&	\sum_{i=1}^j \frac{1}{2}	\Pr(   \fullacceventadapt{i} ) \tracenorm{ \finalstateadapt{i} - \finalstateidealadapt{i} } \\ 
			 	=&	\sum_{\substack{i=1\\ i\in \mathcal{T}_p}}^j \frac{1}{2}	\Pr(   \fullacceventadapt{i} ) \tracenorm{ \finalstateadapt{i} - \finalstateidealadapt{i} } \\
			 	 &\leq 	\sum_{\substack{i=1\\ i\in \mathcal{T}_p}}^j   \Pr(\fullacceventadapt{i} )    \leq 	\sum_{\substack{i=1\\ i\in \mathcal{T}_p}}^j 	\Pr(\Omega_i   ) \\
			 		& \leq \Pr\left( \bstat(\Fobs ) \geq l_j + \leak{j} + \PAcost \right)  \\
			 		&\leq   \Pr\left( \bstat(\Fobs ) \geq H_\alpha(Z^n | C^n E^n)_{\rho}  \right) \leq \epsAT. 
			 	\end{aligned}
			 \end{equation}
		Here the first equality follows from the fact that the real and the ideal outputs are identical when the length of the key generated is zero, the second inequality uses the fact that the trace norm is upper bounded by $2$, and the third inequality follows from the properties of probabilities. 
		The fourth inequality uses \cref{lemma:ordering}, the fifth inequality follows from the fact that  $ l_j + \leak{j} + \PAcost \geq H_\alpha(Z^n | C^n E^n)_{\rho}$, and the final inequality from \cref{lemma:bstat}.
			 
	 For the remaining terms, we follow the same steps as \cref{eq:part2} from the proof of \cref{thm:main}. We obtain the following inequalities:
	 \begin{widetext}
			 \begin{equation} \label{eq:newpart2}
			 	\begin{aligned}
			 		&\sum_{i=j+1}^{M} \frac{1}{2}	\Pr(\fullacceventadapt{i}) \tracenorm{ \finalstateadapt{i} - \finalstateidealadapt{i} }   \\
			 		&=\sum_{\substack{i\geq j+1 \\ i \in \mathcal{T}_p} }^M  \frac{1}{2}	\Pr(\fullacceventadapt{i}) \tracenorm{ \finalstateadapt{i} - \finalstateidealadapt{i} }    \\
			 		&\leq \sum_{\substack{i\geq j+1 \\ i \in \mathcal{T}_p} }^M  \frac{1}{2} \Pr(\fullacceventadapt{i})2^{ -\left(  \frac{\alpha-1}{\alpha}\right)   \left( H_\alpha (Z^n | C^n \Cat^m C_{E} C_V E^N)_{\rho| \fullacceventadapt{i}} - l_i\right) + \frac{2}{\alpha}  - 1     }  \\
			 		&\leq 	\sum_{\substack{i\geq j+1 \\ i \in \mathcal{T}_p} }^M	\frac{1}{2} \Pr( \fullacceventadapt{i} )2^{ -\left(  \frac{\alpha-1}{\alpha}\right)   \left( H_\alpha (Z^n | C^n \Cat^m  C_V E^n)_{\rho| \fullacceventadapt{i} } - \text{leak}_i -  l_i\right) + \frac{2}{\alpha}  - 1     }  \\
			 		&\leq 	\sum_{\substack{i\geq j+1 \\ i \in \mathcal{T}_p} }	^M \frac{1}{2} \Pr(\fullacceventadapt{i} )2^{ -\left(  \frac{\alpha-1}{\alpha}\right)   \left( H_\alpha (Z^n | C^n \Cat^m C_V  E^n)_{\rho| \fullacceventadapt{i}} - \text{leak}_{j+1} -  l_{j+1}  \right) + \frac{2}{\alpha}  -1 }  \\
			 		&\leq 		\frac{1}{2} 2^{ -\left(  \frac{\alpha-1}{\alpha}\right)   \left( H_\alpha (Z^n | C^n C_V E^n)_{\rho} - \text{leak}_{j+1} -  l_{j+1} \right) + \frac{2}{\alpha}    -1 }  \\
			 			&\leq 		\frac{1}{2} 2^{ -\left(  \frac{\alpha-1}{\alpha}\right)   \left( H_\alpha (Z^n | C^n  E^n)_{\rho} - \text{leak}_{j+1} -  l_{j+1}  - \EVCost  \right) + \frac{2}{\alpha}     -1 }  \\
			 			&\leq 		\frac{1}{2} 2^{ -\left(  \frac{\alpha-1}{\alpha}\right)   \left( \PAcost  - \EVCost  \right) + \frac{2}{\alpha} -1      }   \\
			 		& \leq \epsPA.
			 	\end{aligned}
			 \end{equation}
			 \end{widetext}
			 We now explain the derivation of the expressions above, and highlight the crucial steps in \cref{rem:technicalnew}. The first equality follows from the fact that the real and the ideal outputs are identical when the length of the key generated is zero. We use the leftover hashing lemma for \renyi entropy for the second inequality (\cref{lemma:LHLrenyi}), and \cref{lemma:renyisplitting} to split off the information leakage due to error correction, and the $E^{m}$ register (which is independent of $Z^n$) in the third inequality. The ordering on $l_i + \leak{i}$ from \cref{remark:orderingfullyapadative} allows us obtain the fourth inequality, and we use \cref{lemma:renyiweightedaverage} to get rid of the conditioning on events for the fifth inequality. We use \cref{lemma:renyisplitting} again to split off the error-verification communication, and the $\Cat^m$ register (which is independent of $Z^n$) in the sixth inequality. We use the fact that $H_\alpha(Z^n | C^n E^n)_{\rho} \geq l_{j+1} + \leak{j+1} + \PAcost$ for the seventh inequality, and \cref{eq:livalue} for the  eighth inequality.
			 	
			\begin{remark} \label{rem:technicalnew}
				As in the proof of \cref{thm:main}, the critical steps in the above chain of inequalities are the replacement of $l_i+\leak{i}$ with $l_{j+1}+\leak{j+1}$ in the third inequality, and using \cref{lemma:renyiweightedaverage} in the fourth inequality to get rid of the conditioning on events in the \renyi entropies. As in the proof of \cref{thm:main}, we split of $\Cat^m$ and $C_V$ registers  \emph{after} using \cref{lemma:renyiweightedaverage}, since we need the events $\fullacceventadapt{i}$ to be known to Eve in order to use  \cref{lemma:renyiweightedaverage}.
			\end{remark}
			
			\textbf{Case 1 and Case 3:}	 The analysis of Case 1 is a special case of Case 2, and is obtained by setting $j=1$ in Eq.~\eqref{eq:newpart2}. The analysis of Case 3 is a special case of Case 2, and is obtained by setting $j=M$ in Eq.~\eqref{eq:newpart1}.
			
	 Bringing Eqs.~\eqref{eq:newpart1} and ~\eqref{eq:newpart2} together, we obtain
			 \begin{equation}
			 	\begin{aligned}
			 	&\sum_{i=1}^{M}  \frac{1}{2}	\Pr(\fullacceventadapt{i} ) \tracenorm{ \finalstateadapt{i} - \finalstateidealadapt{i} } \\
			 	&\leq  \epsilon_{\text{AT}}+ \epsilon_{\text{PA}} 
			 	\end{aligned}
			 \end{equation}
			 
		 Since the protocol is $\epsEV$-correct, and $(\epsAT+\epsPA)$-secret, it is also $(\epsEV+\epsAT+\epsPA)$-secure. 
			\end{proof}
			\begin{remark}
 Since the protocol from \cref{thm:second} does not impose any condition on the sets $Q_i$, unlike \cref{thm:main} which requires the acceptance sets $\accset{i}$ of the fixed-length protocols to form a nested sequence, one can use \cref{thm:second} in scenarios where the channel behaviour is unpredictable and chaotic. This is especially desirable for ground-to-satellite QKD, where the channel behaviour is difficult to predict in advance. In the next section, we show how \cref{thm:second} can be used to improve the expected key rate  in such scenarios.
			\end{remark}
			\section{Application to Qubit BB84} \label{sec:resultstrueadaptive}
		
				\begin{figure}
				\includegraphics[width=\linewidth]{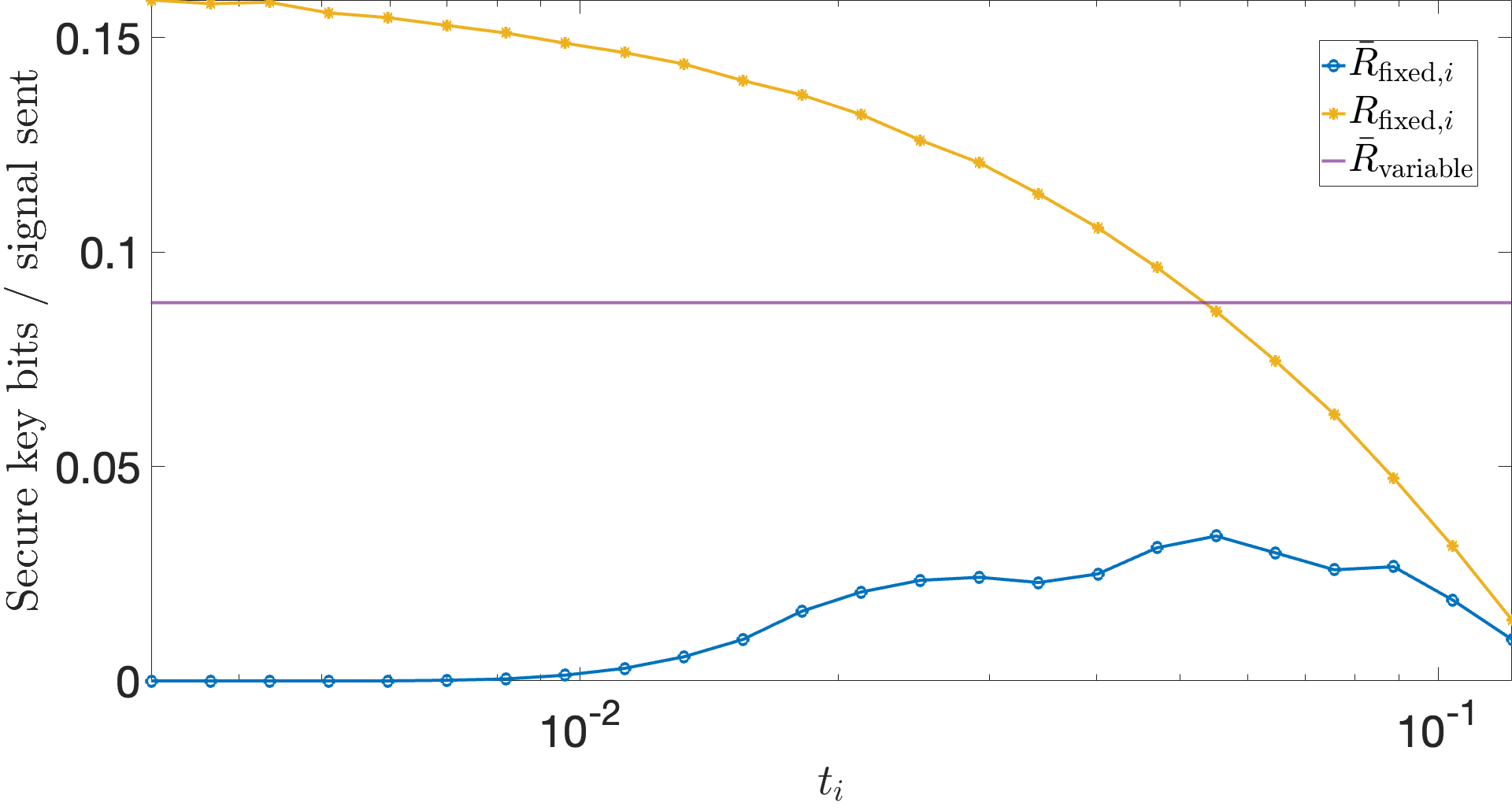} 
				\caption{Expected key rate for fixed-length protocols ($\bar{R}_{\text{fixed},i}$) for various values of $t_i$, key rate upon acceptance for fixed-length protocols ($R_{\text{fixed},i}$) for various values of $t_i$ and the expected key rate for variable-length protocol ($\bar{R}_\text{variable}$).} \label{fig:expkeyratefullyadaptiveplot} 
			\end{figure}
			
	In this section, we compute expected key rates for fixed-length and variable-length qubit BB84 protocols for a scenario where the honest behaviour is unpredictable. The fixed-length implementation is identical to the one from \cref{sec:applicationbb84}. The variable-length implementation is also similar, except the variable-length decision, which is implemented as described in \cref{subsec:protspec} above. In particular, after signal transmission, measurements, and public announcements, Alice and Bob compute $\Fobs$	from public announcements, and determine $l(\Fobs)$ and $\leak{}(\Fobs)$ according to \cref{eq:livalue}.

	For the sake of simplicity, we consider a channel model that can take a discrete set of values for the depolarization probability and the misalignment angle. We assume that the channel is such that, on any given run, the depolarization probability is chosen randomly from $\{ 0.02,0.03,0.04,0.05\}$  with equal probability, and the misalignment angle is chosen randomly from  $\{2^\circ, 4^\circ, 6^\circ, 8^\circ, 10^\circ \}$ with equal probability.  Thus the channel has $\nch = 20$ possible values, which it takes with equal probability. We use  $\honest^{(j)}$ to denote the state corresponding to the $j$th honest behaviour of the channel.  We set the basis choice probabilities to $p_x=p_z=0.5$, the total number of signals to $N=10^6$, and the number of signals used in the public announcement to $m=0.05N$. We estimate the number of bits to be used for error-correction from $\Fobs$, by setting $\leak{}(\Fobs) =f n  H(Z|YC)_{\Fobs}$ where $f=1.16$ is the efficiency factor. We set a target security parameter of $\epsSec = 10^{-12}$. 
		
	We now explain the various key rates plotted in \cref{fig:expkeyratefullyadaptiveplot}. 
		\begin{enumerate}
		\item $\KRfixed{i}$: This is the key rate upon acceptance for the fixed-length protocol, plotted against $t_i$, the size of the acceptance set, and is identical to the plot from \cref{fig:expkeyrateplot}.
		We set $\epsPA=\epsEV=\epsAT=\epsSec/2$, and compute $\KRfixed{i} = l_i/N$, where $l_i$ is computed according to \cref{eq:bb84keyrate} for the acceptance test in \cref{eq:acceptancetest}. We choose the centre ($\Fbar$)  of the fixed-length acceptance set (\cref{eq:acceptancetest}) to be the expected frequency of outcomes corresponding to the channel with the least possible depolarization probability ($0.02$) and least possible misalignment angle ($2^\circ$). As expected, we see that $\KRfixed{i}$ decreases monotonically as we increase $t_i$, reflecting the fact that larger acceptance sets lead to lower key rates upon acceptance.
		\item $\bar{R}_\text{fixed,i}$: This is the expected key rate for fixed-length protocols. Let $\accset{i}$ be the acceptance set for the fixed-length protocol for  a given value of $t_i$. Then, the expected key rate is given by
		\begin{equation}
			\bar{R}_\text{fixed,i}= \frac{1}{\nch}\sum_{j=1}^{\nch}  \left(  \sum_{i=1}^M \Pr( \Fobs \in \accset{i})_{\honest^{(j)}}   \KRfixed{i} \right),
			\end{equation}
			where the expression in the parenthesis represents the expected key rate for the $j$th channel behaviour. We use the values of $\KRfixed{i}$ obtained above, and numerically estimate $ \Pr( \Fobs \in \accset{i})_{\honest^{(j)}}$ from the following steps.
			\begin{enumerate}
				\item For each channel model, we compute the probability vector 	 $\Fbar^{(j)}$ corresponding to the honest behaviour $\honest^{(j)}$, by setting $\Fbar^{(j)}=\Phi(\honest^{(j)})$, where $\Phi$ is defined in \cref{eq:Phimap}.
				\item From each $\Fbar^{(j)}$, we sample $m$ times to obtain $\Fobs$. We check whether $\Fobs \in \Sset{i}$ or not.
				\item We estimate $ \Pr( \Fobs \in \accset{i})_{\honest^{(j)}}$ by repeating step (b) $50$ times for reach possible honest behaviour, and computing the fraction of times we obtained $\Fobs \in \Sset{i}$.
			\end{enumerate}
			Thus, the whole process is a simulation of $1000$ runs of the QKD protocol, with each channel being used $50$ times.
			
			In \cref{fig:expkeyratefullyadaptiveplot} we see that $\bar{R}_\text{fixed,i}$ is much smaller than $\KRfixed{i}$ since the fixed-length protocol only accepts on a small number of channel behaviours.  As $t$ increases, the fixed-length acceptance set becomes larger, starts accepting on multiple values of $\Fbar^j$, and therefore has a larger probability of acceptance. Thus, the expected key rate $\bar{R}_{\text{fixed},i}$ increases slightly.	However, the size of the acceptance test is already large, and the key rate upon acceptance ($\KRfixed{i}$), rapidly goes to zero for large $t_i$. This causes $\bar{R}_\text{fixed,i}$ to also go to zero rapidly.

		\item $\bar{R}_\text{variable}$: This is the expected key rate for variable-length protocols. This is given by
		\begin{equation}
		\bar{R}_\text{variable} = 	 \frac{1}{\nch}\sum_{j=1}^{\nch}  \left( \sum_{\Fobs} \Pr(\Fobs)_{\honest^{(j)}} R_\text{variable}(\Fobs) \right) 
		\end{equation}
		where $\Pr(\Fobs)_{\honest^{(j)}} $ is the probability of obtaining $\Fobs$ when the honest behaviour is $\honest^{(j)}$, and $R_\text{variable}(\Fobs)$ is the key rate obtained for the observed frequency $\Fobs$. The term in the parenthesis represents the expected key rate for the $j$th channel behaviour. Note that $\bar{R}_\text{variable}$ is a fixed value and \textit{not} plotted against $t_i$, and is computed as follows.
		
		\begin{enumerate}
			\item  For each channel model, we compute the expected statistics $\Fbar^{(j)}$ corresponding to the honest behaviour $\honest^{(j)}$.
			\item From each $\Fbar^{(j)}$, we sample $m$ times to obtain $\Fobs$. We compute $l(\Fobs)$ according to \cref{eq:livalue}, with $\epsPA=\epsAT=\epsSec/4$ and $\epsEV=\epsSec/2$, and set $R_\text{variable}(\Fobs) = l(\Fobs)/N$.
			\item For each channel model, we repeat step (b) $50$ times, and compute the average value of $R_\text{variable}(\Fobs) $. This is our estimate of $ \left( \sum_{\Fobs} \Pr(\Fobs)_{\honest^{(j)}} R_\text{variable}(\Fobs) \right) $.
			\item $\bar{R}_\text{variable}$ is then computed by averaging the key rate obtained in step (c), over all the possible channel models.
			\end{enumerate}
			Thus the above procedure is a simulation of $1000$ runs of the QKD protocol, with each channel behaviour being used $50$ times.
			 Crucially, we find the expected key rate for variable-length protocols is much higher than the expected key rate for the fixed-length protocols. 
			\end{enumerate}

	\begin{remark}
		Note that the degree of improvement shown by the variable-length protocol in \cref{fig:expkeyratefullyadaptiveplot} depends on $\nch$. Larger values of $\nch$ reflect a higher variation in the channel behaviour, and will lead to a bigger difference between the performance of fixed-length and variable-length protocols. In this work, we chose the above channel model for the sake of simplicity.  Detailed studies of practical QKD protocols over realistic, unpredictable channel models  will be the subject of future work.
	\end{remark}

	\section{Variable Input-length privacy amplification} \label{sec:variablelengthPA}
So far we have studied  the variable-length aspects of the \textit{final key} that is generated after privacy amplification in QKD protocols. In this section, we will turn our attention to the variable-length aspect of the \textit{sifted raw key} in QKD implementations, before privacy amplification. In particular, we will point out and remedy a gap between the theoretical analysis of privacy amplification and its experimental implementation.
	For simplicity, we only consider fixed-length QKD protocols. However, our results can be generalized to variable-length protocols in a straightforward manner. 
	
	\subsection{Sifting in QKD}
 Consider the following three ways of implementing the sifting step in QKD protocols. 
	
	\begin{enumerate}
		\item\label{PAcase1} \textbf{Map the discard outcomes to $\bot$:} In this case, the state prior to privacy amplification is given by $\rho_{\Zbot^n Y^n \Cbar E^N }$, where $\Zbot$ is a register that takes values in $\{0,1,\bot\}$, and  $\Cbar=C^n\Cat^mC_EC_V$ for brevity. In this case, one has to implement privacy amplification using two-universal hashing from $\hat{Z}^n$ to $l$ bits. In particular, binary Toeplitz hashing, a widely used choice, is not possible.
		\item\label{PAcase2} \textbf{Map the discard outcomes to $0$:} In this case, the state prior to privacy amplification is given by $\rho_{Z^n Y^n \Cbar E^N }$, where $Z$ is a register that takes values in $ \{0,1\}$. In this case, one has to implement privacy amplification using two-universal hashing from $n$ bits to $l$ bits. In particular, binary Toeplitz hashing, a widely used choice, is possible; however, the hash matrices must always be for input strings of a \emph{fixed} length $n$.
		\item\label{PAcase3} \textbf{Actually discard the discard outcomes:}  In this case, the state prior to privacy amplification is given by $\rho_{\Zv Y^n \Cbar E^N }$, where $\Zv $ is a register that takes values in the set of bitstrings of length less than or equal to $n$, which we shall denote as $\{0,1\}^{\leq n}$. In this case, one first looks at the number of bits in the register $\Zv$, denoted by $\mathrm{len}(\Zv)$, and chooses a two-universal hashing procedure from $\mathrm{len}(\Zv)$ bits to $l$ bits. This is what is commonly done in QKD experiments. Practically, one would like to use binary Toeplitz hashing in this procedure. However, we will see below that this is not a valid two-universal hashing procedure from $\{0,1\}^{\leq n}$ to $l$ bits. 
	\end{enumerate}
The theoretical analysis of Case~\ref{PAcase1} and Case~\ref{PAcase2} is straightforward, since they constitute valid two-universal hashing procedures from  $\{0,1,\bot\}^n$ to $l$ bits, and $n$ bits to $l$ bits respectively. Thus, the leftover hashing lemma can be directly applied. However, Case~\ref{PAcase3} is \textit{not}  necessarily a two-universal hashing procedure from $\{0,1\}^{\leq n}$ to $l$ bits, as we now explain. Thus we cannot directly apply the leftover hashing lemma in this case.
	\subsection{The problem} \label{subsec:problem}
	For every $i \in \{0,1,\dots,n\}$, let $\hashfamily{i}$ denote a two-universal hash family from $i$ bits to $l$ bits. Then, the procedure described in Case~\ref{PAcase3} above is equivalent to \textit{first}  randomly sampling $f_i \in \hashfamily{i}$ for every $i$, followed by computing $f_{\mathrm{len}(\Zv)}(\Zv)$. Note that in this case, only one of the sampled $f_i$s is  ever applied. 
		In order for this procedure to be a valid two-universal hashing procedure from $\{0,1\}^{\leq n}$ to $l$, by definition it must be the case that for any two inputs $\zvone \neq \zvtwo$, we have
		\begin{equation} \label{eq:twouniversalhashing}
			\Pr_{f_1,f_2,...,f_n} [f_{\mathrm{len}(\zvone)} (\zvone) = f_{\mathrm{len}(\zvtwo)} (\zvtwo)] \leq \frac{1}{2^l}. 
		\end{equation}
		 When $\zvone$ and $\zvtwo$ are of the same length, then \cref{eq:twouniversalhashing} follows from the two-universal property of $\hashfamily{i}$.
		 When $\zvone$ and $\zvtwo$ are of different length, an explicit counter-example can be obtained by considering $\zvone$ and $\zvtwo$ to be all-zero strings of different lengths. In this case, if $\hashfamily{i}$ is a two-universal \textit{linear} hash family, then $f_{\mathrm{len}(\zvone)} (\zvone) = f_{\mathrm{len}(\zvtwo)} (\zvtwo) = \vecb{0}$ with probability 1. Thus for binary Toeplitz hashing, Case~\ref{PAcase3} is \textit{not} a valid two-universal hashing procedure. Thus we cannot directly apply the leftover hashing lemma. 
		 \begin{remark} \label{remark:hashing}
		 	We note that if every $\hashfamily{i}$ is chosen such that it is two-universal \textit{and} has  the following ``uniform output'' property:
		 	\begin{equation}\label{eq:uniformoutput}
		 		 \Pr_{f_i \in \hashfamily{i}} [f_i(\vecb{z}) = k] \leq 1/2^l \quad \forall \quad  \vecb{z} \in \{0,1\}^i,k\in\{0,1\}^l,
		 	\end{equation}
		 then it is straightforward to prove that \cref{eq:twouniversalhashing} holds and hence the described procedure is a valid two-universal hashing. Furthermore, in principle any two-universal hashing procedure can be modified into one that satisfies \cref{eq:uniformoutput}, via the construction we describe in the \cref{lemma:variableinputhashing} proof below. However, physically implementing this conversion in an actual QKD protocol would be an undesirable additional cost, hence we instead provide a proof that shows that this is not necessary.
		 \end{remark}
		 
		 \subsection{The solution}
		We address this issue with \cref{lemma:variableinputhashing,lemma:renyiinvariance} below. We start by proving the following modified leftover hashing lemma that is applicable to Case~\ref{PAcase3}, as long as the protocol satisfies the property that the positions and values of the discarded outcomes can be determined from the public announcements $\Cbar$ (we return to this point after presenting the lemmas and their proofs). Our approach is to first use \cref{remark:hashing} to construct a virtual hashing procedure that is a valid two-universal hashing procedure from $\{0,1\}^{\leq n} $ to $l$ bits.  We will then show that the actual output states can be obtained by performing a CPTP map on the virtual output states.  The required result then follows from data-processing inequalities. 
		
\newcommand{\rawkey}{R}
\newcommand{\suplabels}{}
\newcommand{\suplabelsbrkt}{}
\begin{lemma} \label{lemma:variableinputhashing}
Let $\rho_{\Zv \Cbar E^N}$ be a state classical in $\Zv \Cbar$ (where the $\Zv$ register takes values in $\{0,1\}^{\leq n}$), with the property that conditioned on each possible value $\bar{c}$ on the $\Cbar$ register, the resulting distribution on $\Zv$ is only supported on values in $\{0,1\}^{k_{\bar{c}}}$ for some constant $k_{\bar{c}} \in \mathbb{N}$.
Let $\rho^{\suplabelsbrkt}_{K_A \Cfull E^N}$ be the state obtained from $\rho_{\Zv \Cbar E^N}$ by first computing the number of bits $\mathrm{len}(\Zv)$ in the $\Zv$ register, then implementing a two-universal hashing procedure from $\mathrm{len}(\Zv)$ bits to $l$ bits, where $\Cfull \coloneqq \Cbar C_P$ with $C_P$ being the choice of hashing function (in other words, the procedure described above in Case~\ref{PAcase3}). Then for any event $\Omega$ on the classical register $\Cbar$, we have (for $\rho^{(\suplabelsbrkt\text{ideal})}_{K_A \Cfull E^N} \coloneqq \frac{\mathbb{I}_{K_A}}{|K_A|} \otimes \rho^{\suplabelsbrkt}_{\Cfull E^N}$):
\begin{equation} \label{eq:varLHL}
\begin{aligned}
&	\half \Pr(\Omega) \tracenorm{ \rho^{\suplabelsbrkt}_{K_A \Cfull E^N | \Omega} - \rho^{(\suplabelsbrkt\text{ideal})}_{K_A \Cfull E^N | \Omega}} \\
&\leq 	\half \Pr(\Omega)  2^{- \left( \frac{\alpha-1}{\alpha}  \right)\left( H_\alpha(\Zv | \Cbar  E^N)_{\rho | \Omega} - l \right) + \frac{2}{\alpha} - 1} \\
& \leq \half 2^{- \left( \frac{\alpha-1}{\alpha}  \right)\left( H_\alpha(\Zv | \Cbar E^N)_{\rho} - l \right) + \frac{2}{\alpha} - 1}  
.
\end{aligned}
\end{equation}

\end{lemma}

\begin{proof}
As explained in \cref{subsec:problem}, the hashing procedure described above can be thought of as \textit{first} randomly sampling $f_i \in \hashfamily{i}$ for every $i$, and then computing $f_{\mathrm{len}(\Zv)}(\Zv)$. However, as noted in that section, this process is not a valid two-universal hashing procedure from $\{0,1\}^{\leq n}$ to $l$ bits.
	
Consider instead the following virtual hashing process, based on new hash families~\footnote{This specification of $\virthashfamily{i}$ is not technically a set of \emph{functions}, since each element of $\virthashfamily{i}$ is instead a tuple where the second term is an $l$-bit string. However, each such element uniquely specifies a function in a simple manner that we shall shortly specify.} $\virthashfamily{i} \coloneqq \hashfamily{i} \times \{0,1\}^l$ (for every $i$). This virtual process first randomly samples $(f_i,u_i) \in \virthashfamily{i}$ for every $i$, i.e.~$f_i$ is sampled from the same two-universal hash family $ \hashfamily{i}$ as before, and $u_i$ is a random $l$-bit string. It then computes $  f_{\mathrm{len}(\Zv)} (\Zv) \oplus u_{\mathrm{len}(\Zv)}$ as its hash output. 
Now, this virtual hashing procedure \textit{is} a valid two-universal hashing procedure from $\{0,1\}^{\leq n}$ to $l$ bits, because each hash family $\virthashfamily{i}$ is two-universal \textit{and} satisfies the ``uniform output'' property (\cref{eq:uniformoutput}). 

Denote the output state of the virtual process (acting on $\rho_{\Zv \Cbar E^N}$) as $\rho^{(\suplabelsbrkt\text{virtual})}_{K_A  \hat{\Cfull}  E^N}$, where $\hat{\Cfull} = \Cbar \hat{C}_P$ with $\hat{C}_P$ being the description of the hash function chosen in the virtual process (in particular, all the values $(f_i, u_i)$ from the virtual process). Let us analogously define $\rho^{(\suplabelsbrkt\text{ideal},\text{virtual})}_{K_A\hat{\Cfull} E^N } \coloneqq \frac{\mathbb{I}_{K_A}}{|K_A|} \otimes \rho^{(\suplabelsbrkt\text{virtual})}_{\hat{\Cfull}  E^N}$.

Now, we construct a CPTP map $\mathcal{E}: K_A\hat{\Cfull} E^N \to K_A \Cfull E^N$ that will map the virtual output states to the actual output states. This map $\mathcal{E}$ does the following operations:
\begin{enumerate}
\item Look at $\Cbar$ and determine the corresponding value $k_{\bar{C}}$ (as defined in the conditions of this lemma)~\footnote{$\mathcal{E}$ cannot ``directly'' compute $\mathrm{len}(\Zv)$ because the register $\Zv$ is no longer present in the states it acts on.}, to be used in the subsequent steps.  
\item Look at $\hat{C}_P$ and determine $u_{k_{\bar{C}}}$, to be used in the subsequent steps.
\item Replace $K_A$ with $K_A \oplus u_{k_{\bar{C}}}$.
\item Partial trace on the $\hat{C}_P$ register, on everything except the $f_{k_{\bar{C}}}$ information.
\end{enumerate}
It is straightforward to verify that this map $\mathcal{E}$ indeed satisfies
\begin{equation}
\begin{gathered}
\mathcal{E} \left( \rho^{(\suplabelsbrkt \text{virtual})}_{K_A \hat{\Cfull} E^N} \right) = \rho^{\suplabelsbrkt}_{K_A \Cfull E^N}, \\
\mathcal{E} \left(  \rho^{(\suplabelsbrkt\text{ideal},\text{virtual})}_{K_A\hat{\Cfull} E^N} \right) = 
\rho^{(\suplabelsbrkt\text{ideal})}_{K_A \Cfull E^N},
\end{gathered}
\end{equation}
and analogously for the above states conditioned on the event $\Omega$ (since $\mathcal{E}$ does not disturb the register $\Cbar$).

Therefore,  we have
\begin{equation}
\begin{aligned}
&	\half \Pr(\Omega) \tracenorm{ \rho^{\suplabelsbrkt}_{K_A \Cfull E^N | \Omega} - \rho^{(\suplabelsbrkt\text{ideal})}_{K_A \Cfull E^N | \Omega}}  \\
& =	\half \Pr(\Omega) \tracenorm{ \mathcal{E} \left( \rho^{(\suplabelsbrkt \text{virtual})}_{K_A \hat{\Cfull} E^N | \Omega} - \rho^{(\suplabelsbrkt\text{ideal},\text{virtual})}_{K_A\hat{\Cfull} E^N | \Omega} \right)} \\
& \leq 	\half \Pr(\Omega) \tracenorm{  \rho^{(\suplabelsbrkt \text{virtual})}_{K_A \hat{\Cfull} E^N | \Omega} - \rho^{(\suplabelsbrkt\text{ideal},\text{virtual})}_{K_A \hat{\Cfull} E^N | \Omega} } \\
&\leq 	\half \Pr(\Omega)  2^{- \left( \frac{\alpha-1}{\alpha}  \right)\left( H_\alpha(\Zv | \Cbar  E^N)_{\rho | \Omega} - l \right) + \frac{2}{\alpha} - 1} \\
&\leq \half 2^{- \left( \frac{\alpha-1}{\alpha}  \right)\left( H_\alpha(\Zv | \Cbar E^N)_{\rho} - l \right) + \frac{2}{\alpha} - 1}
\end{aligned}
\end{equation}
where we used the fact that CPTP maps cannot increase trace norm in the third inequality, and
leftover hashing lemma for \renyi entropies (\cref{lemma:LHLrenyi}) for the fourth inequality, and \cref{lemma:renyiconditioning} for the final inequality.
		\end{proof}
		
			\begin{remark}
			While here we have focused on proving an analogue of the leftover hashing lemma for \renyi entropy (\cref{lemma:LHLrenyi}), a similar result for the smooth min-entropy version can be obtained by exactly the same proof (except that when conditioning on the event $\Omega$, one should use the subnormalized conditional states; see \cite[Lemma~10 and Proposition~9]{tomamichelLargely2017}). 
		\end{remark}
In order to use \cref{lemma:variableinputhashing}, we have to compute bounds on the \renyi entropy $ H_\alpha(\Zv | \Cbar E^N)_{\rho}$, which is computed on the state just prior to privacy amplification in Case~\ref{PAcase3}. However, we expect that if the registers that were discarded to produce $\Zv$ are completely determined by the register $\Cbar$, then this entropy should be the same as the value before the discarding process, since the conditioning register $\Cbar$ could be used to isometrically convert between the values before and after discarding some registers. We formalize this claim in the following Lemma and subsequent discussion.
		
\begin{lemma} \label{lemma:renyiinvariance}
	Suppose $\rho_{\rawkey \Cbar E^N},\rho_{\Zv \Cbar E^N}$ are states that are classical in $\Cbar$, and related to each other as follows: letting $\rawkey_{\bar{c}}$ be a register containing the support of the conditional state $\rho_{\rawkey|\Cbar = \bar{c}}$, there exist isometries $V^{(\bar{c})}_{\rawkey_{\bar{c}} \rightarrow \Zv}$ such that~\footnote{\cref{eq:unitaryaction} is a well-defined expression despite the fact that $V$ is not defined on all of $\rawkey \Cbar$, because $\rho_{\rawkey \Cbar E^N}$ is only supported on the subspace on which $V$ is defined.}
	\begin{equation}  \label{eq:unitaryaction}
		\begin{gathered}
			V\rho_{\rawkey \Cbar E^N} V^\dagger =\rho_{\Zv \Cbar E^N}, \text{ where} \\
			V \coloneqq \sum_{\bar{c}} V^{(\bar{c})}_{\rawkey_{\bar{c}} \rightarrow \Zv} \otimes \ket{\bar{c}}\bra{\bar{c}}_{\Cbar}.
		\end{gathered}
	\end{equation}
	Then we have
	\begin{equation} \label{eq:renyiinvariance}
		H_\alpha(\Zv |\Cbar E^N)_\rho = H_\alpha(\rawkey | \Cbar E^N)_\rho.
	\end{equation}
\end{lemma}

\begin{proof} 
We intuitively expect \cref{eq:renyiinvariance} to be true, since \cref{eq:unitaryaction} essentially states that $\Cbar$ can be used to isometrically convert $\rawkey$ to $\Zv$. 
To formalize this, we first note that each isometry $V^{(\bar{c})}_{\rawkey_{\bar{c}} \rightarrow \Zv}$ can always be extended to an isometry $V^{(\bar{c})}_{\rawkey \rightarrow \Zv}$, i.e.~where the domain is the full Hilbert space of $\rawkey$ (padding the output space $\Zv$ with extra dimensions if $\dim(\rawkey) > \dim(\Zv)$). Furthermore, \cref{eq:unitaryaction} still holds with $V$ defined in terms of these new isometries instead, i.e.~we have
	\begin{equation}  \label{eq:unitaryactiontemp}
	\begin{gathered}
		V\rho_{\rawkey \Cbar E^N} V^\dagger =\rho_{\Zv \Cbar E^N}, \text{ where} \\
		V \coloneqq \sum_{\bar{c}} V^{(\bar{c})}_{\rawkey \rightarrow \Zv} \otimes \ket{\bar{c}}\bra{\bar{c}}_{\Cbar}.
	\end{gathered}
\end{equation}
(It does not matter how we chose the extensions, since $\rho_{\rawkey \Cbar E^N}$ is only supported on a subspace that is unaffected by these choices of extensions.)

Furthermore, letting $\Cbar_c$ be a copy of the register $\Cbar$, using \cite[Lemma B.7]{dupuisEntropy2020} we have 
\begin{equation}
	\begin{aligned}
		H_\alpha(\Zv \Cbar_c | \Cbar E)_\rho &= H_\alpha( \Zv | \Cbar E)_\rho ,\\
		H_\alpha(\rawkey \Cbar_c | \Cbar E)_\rho &= H_\alpha(\rawkey | \Cbar E)_\rho,
	\end{aligned}
\end{equation}
Thus, it is enough to show that $	H_\alpha(\Zv \Cbar_c | \Cbar E)_\rho  = H_\alpha(\rawkey \Cbar_c | \Cbar E)_\rho$. This follows from \cref{eq:unitaryactiontemp}, and the fact that the \renyi entropy is invariant under isometries on the first subsystem, since by defining the isometry $\widetilde{V}_{\rawkey\Cbar_c \rightarrow \Zv \Cbar_c} \coloneqq \sum_{\bar{c}} V^{(\bar{c})}_{\rawkey \rightarrow \Zv} \otimes \ket{\bar{c}}\bra{\bar{c}}_{\Cbar_c}$ we have
\begin{equation}  
	\widetilde{V}\rho_{\rawkey \Cbar_c \Cbar E} \widetilde{V}^\dagger =\rho_{\Zv  \Cbar_c \Cbar E}.
\end{equation}
which concludes the proof \footnote{An alternative proof would be to instead use \cite[Proposition 5.1]{tomamichelQuantum2016} to split the conditional entropies into terms conditioned on each value of $\Cbar$, and note that the equality holds for each term by invariance of \renyi entropy under isometries on the first subsystem.}.
\end{proof}
To apply \cref{lemma:variableinputhashing,lemma:renyiinvariance} in comparing Cases~\ref{PAcase1},~\ref{PAcase2} and~\ref{PAcase3} described previously, we can begin by viewing $\rawkey$ as being $\Zbot^n$ in Case~\ref{PAcase1} or $Z^n$ in Case~\ref{PAcase2}. If the protocol satisfies the condition that the positions and values of discarded outcomes are fixed by the public announcements $\Cbar$, 
we can define operations $V^{(\bar{c})}_{\rawkey_{\bar{c}} \rightarrow \Zv}$ that simply drop the discarded outcomes specified by $\bar{c}$, 
and it is not difficult to show the state $\rho_{\Zv \Cbar E^N}$ in Case~\ref{PAcase3} has the following properties:
\begin{enumerate}
	\item These $V^{(\bar{c})}_{\rawkey_{\bar{c}} \rightarrow \Zv}$ operations are indeed isometries, and $\rho_{\Zv \Cbar E^N}$ is related to $\rho_{\rawkey \Cbar E^N}$ in the sense expressed in \cref{eq:unitaryaction}. 
	
	\item $\rho_{\Zv \Cbar E^N}$ satisfies the conditions of \cref{lemma:variableinputhashing}, and hence \cref{eq:varLHL} is valid.
	
	\item $\rho_{\Zv \Cbar E^N}$ satisfies the conditions of \cref{lemma:renyiinvariance}, and hence $H_\alpha(\Zv |\Cbar E^N)_\rho$ in \cref{eq:varLHL} can be replaced with $H_\alpha(\rawkey |\Cbar E^N)_\rho$.
\end{enumerate}
(Basically, the above statements hold because under that protocol condition, for each value $\bar{c}$, the output length of $V^{(\bar{c})}_{\rawkey_{\bar{c}} \rightarrow \Zv}$ is fixed, and all the discarded positions have fixed values so there are no ``collisions''.)

With this, we see that for Case~\ref{PAcase3} the bound in \cref{eq:varLHL} holds with $H_\alpha(\Zv | \Cbar E^N)_\rho$ replaced by $H_\alpha(\Zbot^n | \Cbar E^N)_\rho$ from Case~\ref{PAcase1} or $H_\alpha(Z^n | \Cbar E^N)_\rho$ from Case~\ref{PAcase2}~\footnote{Note that depending on the proof method used, the bound on $H_\alpha(\Zbot^n |\Cbar E^N)_\rho$ for Case~\ref{PAcase1} may not be equal to the bound on $H_\alpha(Z^n |\Cbar E^N)_\rho$ for Case~\ref{PAcase2}. For instance, with the approach we use in this work, formulas such as \cref{lemma:contrenyi} depend on the dimension of $\Zbot$ versus $Z$. Other proof approaches (for instance, bounding the single-round \renyi entropy directly, rather than first bounding the von Neumann entropy and then applying \cref{lemma:contrenyi}) may not have this feature.}; in particular, for the purposes of this work this means the third line in \cref{eq:newpart2} (and similar bounds in other calculations) is valid even if we apply the procedure in Case~\ref{PAcase3} rather than Case~\ref{PAcase2}.
To qualitatively summarize, under that protocol condition, the bounds obtained on the privacy amplification procedure in QKD are unaffected if the actual protocol implements Case~\ref{PAcase3} in place of Case~\ref{PAcase1} or Case~\ref{PAcase2}. 

	\section{Conclusion} \label{sec:conclusion}
In this work, we presented a security proof for variable-length QKD protocols in the security analysis framework of Renner, against IID collective attacks. First, we showed how a sequence of security proofs for fixed-length protocols satisfying certain conditions can be converted to a security proof for a variable-length protocol. This conversion did not require any new calculations, or any changes to the final key lengths or the lengths of error-correction information. Moreover, the maximum penalty imposed by this approach is a doubling of the security parameter. We exemplified this result by  studying the performance of variable-length and fixed-length implementations of the qubit BB84 protocol, implemented over a fixed, known channel. We showed that the variable-length implementation leads to an improvement in the expected key rate of the protocol, compared to the best fixed-length implementation. Additionally, we showed that implementing the variable-length protocol eliminates the typical trade-off in fixed-length implementations, where a larger acceptance test leads to a higher probability of accepting during honest behaviour, but low key rate upon acceptance.  

Next, we moved on to consider scenarios of unpredictable channels. Here, we construct the variable-length decision in a way that does not rely on a nested sequence of acceptance tests, and proved the security of the resulting class of variable-length protocols.  These protocols did not require users to characterize their channel before running the QKD protocol. Instead, they include instructions for adjusting the length of the final key, and the amount of error-correction information, for every possible observation during the protocol. We exemplified this result by studying the performance of the qubit BB84 protocol implemented in this fashion. We showed that the variable-length implementation leads to a significant improvement in the expected key rate compared to fixed-length implementations, especially for scenarios where the channel is chaotic and unpredictable.

These results are a significant step towards practical QKD implementations, since they eliminate the typical trade-off from fixed-length implementations, and remove the requirement of channel characterization. Moreover,  variable-length protocols have already been implemented in several works based on intuition. This work puts such claims (under the Renner framework) on a solid mathematical footing. 
 (We highlight that in particular, our proof approach relies on a leftover hashing lemma for \renyi entropies that was only recently developed, in Ref.~\cite{dupuisPrivacy2022}. It does not seem entirely straightforward to construct a similar rigorous analysis using the earlier leftover hashing lemma versions that were based on smooth min-entropy.)

In order to use the results of this work to implement a valid variable-length QKD protocol, one can  follow the following steps. First, decide $\leak{}(\Fobs)$ to be \textit{any} function of the observed statistics $\Fobs$. This fixes the number of bits used for error-correction, for any $\Fobs$.  Second, construct a set $V(\Fobs)$ satisfying \cref{eq:VFobscondition}. This fixes $\bstat(\Fobs)$ via \cref{lemma:bstat}, and $l(\Fobs)$ via \cref{eq:livalue}. Then, the variable-length protocol that produces a key of length $l(\Fobs)$, and uses $\leak{}(\Fobs)$ bits for error-correction, upon obtaining $\Fobs$, is secure. In practise, one should choose $\leak{}(\Fobs)$ such that the  error-correction protocol has a high chance of succeeding. 
	Furthermore, while we have provided one construction of $V(\Fobs)$ in this work, it is straightforward to construct $V(\Fobs)$ using other concentration inequalities.

Finally, all our security proofs can be lifted to coherent attacks using the postselection technique. For pedagogical reasons, this is included in Ref.~\cite{nahar2024postselection}, where we also fix a technical flaw in the application of postselection technique to QKD. Alternatively, we highlight that the only part of our \cref{sec:truevariablelength} proof that relied on the IID collective-attacks assumption was the construction of $\bstat$ in \cref{lemma:truevariable,lemma:bstat}. Therefore, any alternative approach that could construct a valid $\bstat$ for coherent attacks would also serve to yield a security proof against such attacks for variable-length protocols.

	\section*{Acknowledgements}
	We would like to acknowledge useful discussions with Renato Renner, especially for most results in \cref{sec:variablelengthPA}. We would like to thank Lars Kamin for helpful discussions on the finite-size security proof of QKD protocols.  We would like to thank John Burniston for help with debugging code. This work was funded by the NSERC Discovery Grant, and was conducted at the Institute for Quantum Computing, University of Waterloo, which is funded by Government of Canada through ISED. This work was partially funded by the Mike and Ophelia Laziridis Fellowship. 	
	
	\bibliography{bibliography_adaptive}

	\appendix 
	
	\section{Technical Definitions and Lemmas}
	\label{appendix:technical}
	We use $S_\circ (A)$ denote the normalized states on $A$. We start by defining the \renyi entropy used in this work.
	
	\begin{definition}[\renyi entropy] \label{def:renyientropy}
		For $\rho \in S_\circ(AB)$, and $\alpha \in (0,1) \cup (1,\infty)$, the sandwiched \renyi entropy of $A$ given $B$ for a state $\rho_{AB}$ is given by
		\begin{equation}
			H_\alpha (A|B)_\rho \coloneqq \max_{\sigma_B \in S_\circ(B)} H_\alpha(A|B)_{\rho | \sigma}	,
				\end{equation}
		where
		\begin{equation}
			\begin{aligned}
				&H_\alpha(A|B)_{\rho| \sigma} \coloneqq \\
				& \begin{cases*}
				\frac{1}{1-\alpha} \log \Tr \left[   \left( \sigma_B^{\frac{1-\alpha}{2\alpha}} \rho_{AB}  \sigma_B^{\frac{1-\alpha}{2\alpha}} \right)^\alpha \right] & if $\rho \in A \otimes \text{supp}(\sigma)$ ,\\
					-\infty     & otherwise.
				\end{cases*}
			\end{aligned}
		\end{equation}
	\end{definition}
	We will require several results regarding the \renyi entropy defined above from \cite{dupuisPrivacy2022, dupuisEntropy2020, tomamichelQuantum2016}.
	Note that the sandwiched \renyi Entropy is referred to as $H_\alpha(A|B)$ in Ref.~\cite{dupuisPrivacy2022} (Definition 1), and $\widetilde{H}^{\uparrow}_{A|B}$ in Ref.~\cite{tomamichelQuantum2016} (Definition 5.2), and $H^{\uparrow}_\alpha(A|B)$ in Ref.~\cite{dupuisEntropy2020} (Definition B.1).  
	
	\begin{lemma}(Leftover hashing lemma using \renyi Entropy \cite[Theorem 8]{dupuisPrivacy2022}) \label{lemma:LHLrenyi}
		Let $\rho_{XE} \in S_\circ(XE)$ be classical on $X$. Let $\mathcal{F}$ be a family of two-universal hash functions from $X$ to $K=\{0,1\}^l$. Let $\omega_K = \sum_{k=1}^{2^l-1} \frac{\ket{k} \bra{k}}{2^l}$ be the perfectly mixed state on $K$, and let $\omega_F = \sum_{f \in \mathcal{F}} \frac{\ket{f} \bra{f} }{|\mathcal{F}|} $.  $\rho_{KEF}$ be the state obtained from $\rho_{XE} \otimes \omega_F$ by applying the two-universal hash function in the register $F$ to $X$. Then, 
		\begin{equation}
			\tracenorm{\rho_{KEF} - \omega_{K} \otimes \rho_{E} \otimes \omega_{F}} \leq 2^{-\frac{(\alpha-1)}{\alpha} \left(     H_\alpha(X|E)_\rho - l    \right) + \frac{2}{\alpha} - 1},
		\end{equation}
		where $\alpha \in (1,2)$.
			\end{lemma} 

	The following two lemmas are used to convert the \renyi entropy of an IID state to the von Neumann entropy on a single round state.

	\begin{lemma}(Additivity of \renyi Entropy, \cite[Corollary 5.2]{tomamichelQuantum2016} ) \label{lemma:additivity}
		For any two states $\rho_{AB} \in S_\circ(AB), \sigma_{CD} \in S_\circ(CD)$, and $\alpha \geq \frac{1}{2}$, we have
		\begin{equation}
			H_\alpha(AC | BD)_{\rho \otimes \sigma}  = H_\alpha(A|B)_\rho + H_\alpha(C|D)_\sigma.
			\end{equation}
	
	\end{lemma}

	\begin{lemma}(\cite[Lemma B.9]{dupuisEntropy2020})\label{lemma:contrenyi}
		For any $\rho_{AB}\in S_\circ (AB)$, and $1 < \alpha < 1+1/ \log(1+2d_A)$, we have
		\begin{equation}
			H_\alpha(A|B)_\rho >  H(A|B)_\rho - (\alpha-1)\log^2(1+2d_A).
		\end{equation}
	\end{lemma}
	In some cases the slightly more elaborate continuity bound derived in \cite[Corollary IV 2]{Dupuis_2019Improved} may perform better than \cref{lemma:contrenyi}; we leave a more detailed analysis for future work.

	\begin{lemma}(Conditioning on Events, \cite[Lemma B.5]{dupuisEntropy2020}) \label{lemma:renyiconditioning}
	Let $\rho_{AB} \in S_\circ(AB)$ be a state of the form $\rho_{AB}= \sum_x p_x \rho_{AB|x}$, where $p_x$ is a probability distribution. Then, for $\alpha>1$, 
				\begin{equation}
					H_\alpha (A|B)_{\rho | x} \geq H_\alpha (A|B)_\rho - \frac{\alpha}{\alpha-1} \log \left( \frac{1}{p_x} \right)	.			
					\end{equation}
	\end{lemma} 
	
	The following Lemma plays a crucial role in getting rid of terms involving the \renyi entropy evaluated on states conditioned on events, in the proofs of \cref{thm:main,thm:second}.
	
	\begin{lemma}  \label{lemma:renyiweightedaverage}
		Let $\rho_{ABCY} = \sum_{y \in \Lambda} p(y) \rho_{ABC| y} \otimes \ket{y} \bra{y} \in S_\circ(ABCY)$ be classical in $Y,C$, where $p(y)$ is a probability distribution over $\Lambda$, and $Y$ can be generated from $C$ (more precisely: $Y \leftrightarrow C \leftrightarrow AB$ forms a Markov chain).  Let $\Lambda^\prime \subseteq \Lambda$. Then, 
		\begin{equation}
			  \sum_{y \in \Lambda^\prime}  p(y)	2^{-\frac{(\alpha-1)}{\alpha} H_\alpha( A| BC)_{\rho_{ |y} } } \leq 2^{-\frac{(\alpha-1)}{\alpha} H_\alpha( A| BC)_\rho } .
		\end{equation}
	\end{lemma} 
	\begin{proof} We have
		\begin{equation}
			\sum_{y \in \Lambda^\prime}  p(y)	2^{-\frac{(\alpha-1)}{\alpha} H_\alpha( A| BC)_{\rho_{ |y} } } \leq \sum_{y \in \Lambda}  p(y)	2^{-\frac{(\alpha-1)}{\alpha} H_\alpha( A| BC)_{\rho_{ |y} } },
		\end{equation}
		since we only add positive terms to the expression to go from the LHS to the RHS. Now, on the RHS, $p(y)$ is a normalized probability distribution function over $\Lambda$. Therefore, we can directly use \cite[Proposition 5.1]{tomamichelQuantum2016}, and we obtain
			\begin{equation}
			\sum_{y \in \Lambda}  p(y)	2^{-\frac{(\alpha-1)}{\alpha} H_\alpha( A| BC)_{\rho_{ |y} } } =	2^{-\frac{(\alpha-1)}{\alpha} H_\alpha( A| BCY)_{\rho}  }.
		\end{equation}
		Since $Y$ can be generated from $C$,  the fact that $H_\alpha( A| BCY) = H_\alpha(A|BC)$ follows by applying the data-processing inequality for \renyi entropy  \cite[Corollary 5.1]{tomamichelQuantum2016} in both directions $YC \rightarrow C$ and $C \rightarrow YC$. Therefore, the claim follows.
	\end{proof}
	
	The following Lemma is used to split off the information leakage due to error-correction and error-verification.
	\begin{lemma}(Splitting off a classical register \cite{tomamichelQuantum2016}) \label{lemma:renyisplitting}
		Let $\rho_{ABCC^\prime} \in S_\circ(ABCC^\prime) = \rho_{ABC} \otimes \rho_{C^\prime}$ be classical on $CC^\prime$, Then, 
		\begin{equation}
			\begin{aligned}
			H_\alpha(A|BCC^\prime)_ \rho&= H_\alpha(A|BC)_\rho \\
			&\geq
			H_\alpha(AC|B)_\rho - \log(\dim(C)) \\
			&\geq  H_\alpha(A|B)_\rho  - \log(\dim(C)) .
			\end{aligned}
			\end{equation}		
	\end{lemma}
	\begin{proof}
		The equality follows from the use of data-processing inequalities \cite[Eq. 5.40]{tomamichelQuantum2016}.
		The inequalities follow from  \cite[Eq. 5.94]{tomamichelQuantum2016}.
	\end{proof}

	Although the fact that $\epscorrect$-correctness and $\epssecret$-secrecy implies $(\epscorrect+\epssecret)$-security for QKD protocols has been shown in many places for fixed-length protocols \cite{ben-orUniversal2004, rennerSecurity2005, portmannCryptographic2014a,portmannSecurity2022}, here we show that the same claim holds for variable-length protocols as well.
	\begin{lemma}[Correctness and Secrecy imply Security] \label{lemma:correctandsecret}
		Consider an variable-length QKD protocol that only produces a key if $\acceventEV$ occurs (error-verification passes). Suppose that the protocol satisfies the correctness condition (\cref{eq:corrdef}):
	 \begin{equation}
			\Pr(K_A\neq K_B \wedge \Omega_{\text{EV}}) \leq \epscorrect,
		\end{equation}
	 and the secrecy condition (\cref{eq:secretdefvar}):
		\begin{equation}
			\sum_{k} \frac{1}{2}	\Pr(\Omega_{\mathrm{len}=k} ) \tracenorm{ \rho^{(k)}_{K_A  \bar{C} E^N| \Omega_{\mathrm{len}=k} } - \rho^{(k,\text{ideal})}_{K_A  \bar{C} E^N| \Omega_{\mathrm{len}=k} } } \leq \epssecret,
		\end{equation}
			where $\Omega_{\mathrm{len}=k}$ is the event that a key of length $k$ is produced.  
		Then the protocol satisfies the security statement
		\begin{equation} \label{eq:secrecystatement}
			\begin{aligned}
			&\sum_{k} \frac{1}{2}	\Pr(\Omega_{\mathrm{len}=k} ) \tracenorm{ \rho^{(k)}_{K_A K_B \bar{C} E^N| \Omega_{\mathrm{len}=k} } - \rho^{(k,\text{ideal})}_{K_A K_B \bar{C} E^N| \Omega_{\mathrm{len}=k} } }\\
			& \leq \epscorrect+\epssecret.
			\end{aligned}
		\end{equation}
	
	\end{lemma}
	\begin{proof}
		Let 
		\begin{equation} \label{eq:statedefs}
			\begin{aligned}
				\rho^{(k)}_{K_A K_B \bar{C} E^N| \Omega_{\mathrm{len}=k}}  &= \sum_{s_A, s_B  \in \{0,1\}^k   } \Pr(s_A,s_B | \Omega_{\mathrm{len} =k}) \\
				&\otimes \ket{s_A,s_B } \bra{s_A,s_B} \otimes \rho^{(k,s_A,s_B)}_{\bar{C}E^N} \\
				\rho^{(k, \text{equal})}_{K_A K_B \bar{C} E^N| \Omega_{\mathrm{len}=k}} & = \sum_{s_A,s_B \in \{0,1\}^k   } \Pr(s_A, s_B | \Omega_{\mathrm{len}=k}) \\
				&\otimes  \ket{s_A,s_A} \bra{s_A,s_A} \otimes \rho^{(k,s_A,s_B)}_{\bar{C}E^N} \\
					\rho^{(k, \text{ideal})}_{K_A K_B \bar{C} E^N| \Omega_{\mathrm{len}=k}} & = \sum_{s_A \in \{0,1\}^k   } \frac{1}{2^k} \ket{s_A,s_A} \bra{s_A,s_A} \otimes \rho^{(k)}_{\bar{C}E^N}.
			\end{aligned}
		\end{equation}
		By the triangle inequality, we have
		
		\begin{equation} \label{eq:splitupsecdef}
			\begin{aligned}
			&\sum_{k} \frac{1}{2}	\Pr(\Omega_{\mathrm{len}=k} ) \tracenorm{ \rho^{(k)}_{K_A K_B \bar{C} E^N| \Omega_{\mathrm{len}=k} } \!- \rho^{(k,\text{ideal})}_{K_A K_B \bar{C} E^N| \Omega_{\mathrm{len}=k} } } \\
			&= 	\sum_{k} \frac{1}{2}	\Pr(\Omega_{\mathrm{len}=k} ) \tracenorm{ \rho^{(k)}_{K_A K_B \bar{C} E^N| \Omega_{\mathrm{len}=k} } \!- \rho^{(k,\text{equal})}_{K_A K_B \bar{C} E^N| \Omega_{\mathrm{len}=k} } } \\
			&+	\sum_{k} \frac{1}{2}	\Pr(\Omega_{\mathrm{len}=k} ) \tracenorm{ \rho^{(k,\text{equal})}_{K_A K_B \bar{C} E^N| \Omega_{\mathrm{len}=k} } \! - \rho^{(k,\text{ideal})}_{K_A K_B \bar{C} E^N| \Omega_{\mathrm{len}=k} } }.  \\
		\end{aligned}
		\end{equation}
We will now relate the first term on the RHS with the correctness condition, and the second term on the RHS with the secrecy condition.
	To bound the first term in Eq.~\eqref{eq:splitupsecdef} we first obtain
	\begin{equation}
		\begin{aligned}
			&\frac{1}{2} \tracenorm{ \rho^{(k)}_{K_A K_B \bar{C} E^N| \Omega_{\mathrm{len}=k} } - \rho^{(k,\text{equal})}_{K_A K_B \bar{C} E^N| \Omega_{\mathrm{len}=k} } } \\
			&\leq \sum_{s_A \neq s_B}  \Pr(s_A, s_B | \Omega_{\mathrm{len}=k})  \\
			& = \Pr(K_A \neq K_B  | \Omega_{\mathrm{len}=k}) ,
		\end{aligned}
	\end{equation}
	where the first inequality follows from the definition of the states in Eq.~\eqref{eq:statedefs}. Therefore,  the first term in Eq.~\eqref{eq:splitupsecdef} can be bounded via
	
	\begin{equation}
		\begin{aligned}
			&\sum_{k} \frac{1}{2}	\Pr(\Omega_{\mathrm{len}=k} ) \tracenorm{ \rho^{(k)}_{K_A K_B \bar{C} E^N| \Omega_{\mathrm{len}=k} } - \rho^{(k,\text{equal})}_{K_A K_B \bar{C} E^N| \Omega_{\mathrm{len}=k} } }  \\
			&\leq \sum_{k} \frac{1}{2} 	\Pr(\Omega_{\mathrm{len}=k} )  \Pr(K_A \neq K_B| \Omega_{\mathrm{len}=k})			 \\
			&=\sum_{k} \Pr(K_A \neq K_B \wedge \Omega_{\mathrm{len} = k} ) \\
			&\leq \Pr(K_A \neq K_B \wedge \Omega_{\text{EV}}) \leq \epscorrect,
		\end{aligned}
	\end{equation}
	where in the penultimate step, we use the fact that a key is produced only if event $\Omega_{\text{EV}}$ occurs.
	
	The second term in Eq.~\eqref{eq:splitupsecdef} is identical to the secrecy statement Eq.~\eqref{eq:secrecystatement} since $K_A = K_B$, and hence we obtain the desired result.
	\end{proof}
	\section{BB84  protocol} \label{appendix:bb84}
  Our protocol is such that in every round Alice and Bob select their basis independently, with probabilities $p_z$ (for $Z$ basis) and $p_x=1-p_z$ (for $X$ basis). If basis $Z$ is selected, Alice sends the states $\ket{0},\ket{1}$ with equal probability. If basis $X$ is selected, Alice sends the states $\ket{+},\ket{-}$ with equal probability. 	Using the source-replacement scheme \cite{curtyEntanglement2004}, this process can be equivalently described as Alice creating the Bell-state $\ket{\psi}_{AA^\prime} = \ket{\phi_+} = \frac{\ket{00}+\ket{11}}{\sqrt{2}}$, and sending $A^\prime$ to Bob. 
  Eve then interacts with the $A^\prime$ system, and forwards the system $B$ to Bob.
  This is then followed by Alice and Bob measuring their respective systems using the POVMS  $\{ P_{(Z,0)} = p_z \ket{0} \bra{0}, P_{(Z,1)} = p_z \ket{1}\bra{1}, P_{(X,0)} = p_x \ket{+} \bra{+} , P_{(X,1)} = p_x \ket{-} \bra{-} \}$. The rest of the protocol steps (such as sifting, key map etc) are the same as in \cref{sec:notation}.   After $n$ such rounds have been performed, Alice and Bob choose a uniformly random subset of size $m$ out of the $n$ rounds, to be publicly announced~\footnote{This procedure is not entirely optimal, since for instance the number of announced rounds where both parties chose the $X$ basis (which is the only ``useful'' data for constraining the entropy of Alice's $Z$-basis rounds) is only approximately $\sim p_x^2 m$. By using a different procedure for choosing the announced rounds, this could be increased to approximately $\sim p_x^2 n$ (see e.g.~\cite{dupuisEntropy2020}~Section~5.1); however, we leave the details for future work.}.
   
   To simulate the channel statistics, we model misalignment as a rotation of angle $\theta$ about the $Y$ axis on $A^\prime$, with 
  \begin{equation}
  	\begin{aligned}
  		U(\theta) &= I_A \otimes \begin{pmatrix}
  			\cos(\theta) & -\sin(\theta) \\
  			\sin(\theta) & \cos(\theta)
  		\end{pmatrix}, \\
  		\mathcal{E}_{\text{misalign}} (\rho) &= U(\theta) \rho U(\theta)^\dagger.
  	\end{aligned}
  \end{equation}
  Depolarization is modelled as a map 
  \begin{equation}
  	\mathcal{E}_\text{depol} (\rho)= (1-q) (\rho)+q \text{Tr}_{A^\prime} (\rho) \otimes \frac{I_B}{2} ,
  \end{equation}
  where $q$ is the depolarization probability. The state on which expected statistics are computed is given by $\honest =  	\mathcal{E}_\text{depol} ( \mathcal{E}_{\text{misalign}}(\ket{\phi_+} \bra{\phi_+}) )$.

		\subsection{Concentration Inequality} \label{appendixsubsec:acceptancetest}
		We state the following lemma from Ref.~\cite{georgeNumerical2021}, which forms the basis of our acceptance tests in the fixed-length protocols and the variable-length decision in the variable-length protocol (\cref{lemma:bstat}, \cref{eq:bb84Sset}).
		\begin{lemma} \label{lemma:concentration}
			Let $\Fbar \in \mathcal{P}(\Sigma)$ be a probability distribution, and let $\Fobs \in \mathcal{P}(\Sigma)$ be a frequency of outcomes obtained from $m$ IID samples from $\Fbar$. Let $\mu=\sqrt{2} \sqrt{\frac{\ln(1/\epsAT) + |\Sigma| \ln(m+1)}{m}}$. Then,
			\begin{equation}
				\Pr( \norm{\Fobs - \Fbar}_1 \geq \mu ) \leq \epsAT
			\end{equation}
		\end{lemma}
		
		\begin{proof}
			From \cite[Theorem 11.2.1 (Sanov's theorem)]{coverELEMENTS}, 
			we obtain 
			\begin{equation} \label{eq:temp1}
				\Pr (D(\Fobs || \Fbar ) > \epsilon)  \leq 2^{-m \left(\epsilon - | \Sigma| \frac{\log_2(m+1)}{m} \right)}
			\end{equation}
			where $D$ is the classical relative entropy. Furthermore, from \cite[Theorem 11.6.1]{coverELEMENTS}, we have
			\begin{equation} \label{eq:temp2}
				\sqrt{2 \ln(2) D(\Fobs || \Fbar)} \geq \norm{\Fobs - \Fbar}_1.
			\end{equation}
			Combining \cref{eq:temp1,eq:temp2} we obtain
			\begin{equation}
				\begin{aligned}
					\Pr ( \norm{\Fobs - \Fbar}_1 \geq \mu) &\leq  \Pr( 	\sqrt{2 \ln(2) D(\Fobs || \Fbar)}  \geq \mu) \\
					&= \Pr (D(\Fobs || \Fbar) \geq \frac{ \mu^2 }{ 2 \ln(2) } ) \\
					& \leq 2^{-m \left( \frac{ \mu^2 }{ 2 \ln(2) }- | \Sigma| \frac{\log_2(m+1)}{m} \right)} .
				\end{aligned}
			\end{equation}
			The required result is obtained by setting $\epsAT = 2^{-m \left( \frac{ \mu^2 }{ 2 \ln(2) }- | \Sigma| \frac{\log_2(m+1)}{m} \right)}$.
		\end{proof}
	\subsection{Numerics} \label{appendixsubsec:krausoperators}
	We use the numerical framework from \cite{winickReliable2018} to compute key rates in this work. This framework equivalently describes the steps in the QKD protocol via Kraus operators $\{K_i\}$, which represent measurements, announcements and sifting done by Alice and Bob, and  $\{Z_j\}$ which implement a pinching channel on the key register. The optimization problem $	\min_{\rho \in S}	 H(Z|CE)_\rho$ is then restated as
	\begin{equation}
		\min_{\rho \in S}	 H(Z|CE)_\rho =  \min_{\rho \in S}	 f(\rho),
	\end{equation}
	where 
	\begin{equation}
		\begin{aligned}
			f(\rho) &= D (\mathcal{G} (\rho) || \mathcal{Z} (\mathcal{G}(\rho) ) ), \\
			\mathcal{G}(\rho) &= \sum_i K_i \rho K_i^\dagger, \\
			\mathcal{Z} (\mathcal{G}(\rho) ) &= \sum_j Z_j \mathcal{G} (\rho) Z_j^\dagger,
		\end{aligned}
	\end{equation}
	and where $D(X || Y) = \text{Tr} (X \log (X) ) - \text{Tr} (X \log (Y))$ is the quantum relative entropy where $\log$ denotes the matrix logarithm. 
	
	The construction of the Kraus operators $K_i$ and $Z_i$ is specified in \cite{winickReliable2018} along with improvements in \cite{linAsymptotic2019}. The Kraus operators for qubit BB84 protocol are given by
	\begin{equation}
		\begin{aligned}
			K_Z &= \left[ \begin{pmatrix} 1 \\ 0 \end{pmatrix}_Z  \otimes  \sqrt{p_z} \begin{pmatrix} 1 & \\ & 0 \end{pmatrix}_A + \begin{pmatrix} 0 \\ 1 \end{pmatrix}_Z \otimes  \sqrt{p_z} \begin{pmatrix} 0 &  \\ & 1  \end{pmatrix}_A \right] \\
			& \otimes \sqrt{p_z} \begin{pmatrix} 1 & \\ & 1 \end{pmatrix}_B \otimes \begin{pmatrix} 1 \\ 0 \end{pmatrix}_{C}, \\
			K_X &= \left[ \begin{pmatrix} 1 \\ 0 \end{pmatrix}_Z \!  \otimes \sqrt{\frac{p_x}{2} } \begin{pmatrix} 1 & 1 \\ 1 & 1 \end{pmatrix}_A \! + \begin{pmatrix} 0 \\ 1 \end{pmatrix}_Z \!   \otimes\sqrt{ \frac{p_x}{2} } \begin{pmatrix} 1 & -1 \\ -1 & 1 \end{pmatrix}_A  \! \right] \\
			& \otimes \sqrt{p_x} \begin{pmatrix} 1 & \\ & 1 \end{pmatrix}_B \otimes \begin{pmatrix} 0 \\ 1 \end{pmatrix}_{C}, 
		\end{aligned}
	\end{equation}
	
	and
	\begin{equation}
		\begin{aligned}
			Z_1 &= \begin{pmatrix} 1 & \\ & 0 \end{pmatrix} \otimes \mathbb{I}_{\dim(A) \times\dim(B) \times \dim(C)}, \\
			Z_2 &= \begin{pmatrix} 0 & \\ & 1 \end{pmatrix} \otimes \mathbb{I}_{\dim(A) \times\dim(B) \times \dim(C)}.
		\end{aligned}
	\end{equation}

\end{document}